\documentclass[%
 reprint,
superscriptaddress,
 amsmath,amssymb,
 aps,
]{revtex4-2}

\usepackage[utf8]{inputenc}				

\usepackage{lmodern} 
\usepackage{microtype}
\usepackage{natbib}
\setcitestyle{numbers,square,sort&compress}
\usepackage{amsmath,amssymb,amsfonts}
\usepackage{algorithmic}
\usepackage{graphicx}
\usepackage{textcomp}
\usepackage{lipsum,mathtools,cuted}
\usepackage{xcolor}

\usepackage{float}
\usepackage{epsfig}
\usepackage{epstopdf}
\usepackage{subfigure,color}
\usepackage{dcolumn}
\usepackage{bm}
\usepackage{lineno}
\usepackage{accents}
\usepackage{multirow}
\usepackage[usestackEOL]{stackengine}
\usepackage{tabularx}
\usepackage{makecell}
\usepackage{dcolumn}
\usepackage{bm}
\usepackage{booktabs}
\UseRawInputEncoding

\usepackage{mdframed}

\begin{document}

\title{Loss-induced performance limits of all-dielectric metasurfaces for terahertz sensing}
\author{J. A. \'Alvarez-Sanchis}
\affiliation{
Nanophotonics Technology Center, Universitat Polit\`cnica de Val\`encia, Valencia 46022, Spain
}
 \email{jaalvsa1@ntc.upv.es}
 
\author{B. Vidal}%
\affiliation{
Nanophotonics Technology Center, Universitat Polit\`cnica de Val\`encia, Valencia 46022, Spain
}

 \email{bvidal@dcom.upv.es}
 
 \author{S. A. Tretyakov}
 \affiliation{
Department of Electronics and Nanoengineering, School of Electrical Engineering, Aalto University, 02150 Espoo, Finland
}

 \author{A. D\'iaz-Rubio}
\affiliation{
Nanophotonics Technology Center, Universitat Polit\`cnica de Val\`encia, Valencia 46022, Spain
}
\email{andiaru@upv.es}

\date{\today}

	
\begin{abstract}
  Metasurfaces providing resonances arising from quasi-bound states in the continuum have been proposed as sensors in the THz band due to the existence of strong resonances characterized by high quality factors.
Controlling geometrical parameters, the quality factor can be adjusted and, theoretically, designed at will. 
However, losses in materials critically bound the metasurface performance and limit the quality factor of the resonances.
For this reason, all-dielectric metasurfaces have been proposed as an alternative to metal-dielectric structures to reduce losses and achieve extreme functionalities. 
When implemented by low-loss materials, these structures are usually considered lossless and proposed as ultrasensitive sensors in the THz band.  
In this paper, we examine the effect of losses in all-dielectric metasurfaces considering realistic materials and study the limitations in the quality factor.
In addition, we compare the performance of these structures as sensors with a nanostructure supporting extraordinary optical transmission.
Our results show that material loss, even in low-loss materials, severely limits the sensing performance in all-dielectric metasurfaces, and that their performance can be surpassed by structures supporting extraordinary optical transmission.  
\end{abstract}


\maketitle

\section{Introduction} 
The THz band, spanning from 100 GHz to 30 THz, has recently received much attention in different research fields \citep{Tonouchi2007cutting} (from medical science  \citep{Woodward2002terahertz} and biology \citep{Fischer2005terahertz} to industrial non-destructive evaluation \citep{NTC_THz}) due to its complementarity to other regions of the spectrum for addressing challenges in the field of sensing. Unfortunately, the sensitivity of THz sensors needs to be boosted to allow a real impact as a solution to industry and clinical demands. One of the main ways of obtaining better sensitivities is the use of metasurfaces.
Metasurfaces are artificial surfaces composed of unit cells, also called meta-atoms, that allow controlling abrupt changes in the amplitude and phase of the transmitted and reflected waves. This control of the electromagnetic (EM) field can be used to control wave scattering \citep{Asadchy2016Perfect}, maximize the absorption by thin layers \citep{Diaz2014Extraordinary}, or produce strong resonances with high field concentrations enhancing light-matter interaction \citep{Achouri,Beruete2020terahertz}. For THz sensors where we need to maximize the interaction with the substance under test, one must rely on strongly resonant metasurfaces with the potential to be exploited for implementing ultrasensitive sensors. Different metasurface-based solutions to increase the capabilities of THz sensors are emerging all along the electromagnetic spectrum \citep{Beruete2020terahertz, Tittl2018Imaging, Leitis2019Angle, Driscoll2007tuned,Hu2015study, Hong2018enhanced, Singh2014ultrasensitive, Yahiaoui2016terahertz,Jauregui2018thz}.

In the literature, several strategies for magnifying the capability of sensors to detect small changes in the substance under test based on metasurfaces can be found, starting from classical split-ring resonator (SRR) arrays \citep{Driscoll2007tuned,Hu2015study},  plasmonic resonances \citep{Hong2018enhanced}, or Fano \citep{Singh2014ultrasensitive} and extraordinary transmission resonances \citep{Yahiaoui2016terahertz,Jauregui2018thz}. Recently, alternative phenomena for enhancing the field concentration have been proposed based on the so-called quasi-bound states in the continuum (quasi-BIC). 
This type of resonance appears when small asymmetries are introduced in the unit cell of a metasurface, transforming trapped modes with an infinite quality factor that could not be excited into resonant modes with a very high quality factor \citep{Fedotov2007sharp}.
Tailoring the geometrical parameters and the symmetry of these structures, one can control the sensor resonant properties (resonant frequency and bandwidth) and use them to study the sample spectral absorption properties. 
\textcolor{black}{Metasurfaces supporting this kind of resonances have been fabricated and experimentally characterized \citep{Liu2021Quasi, Liu2021Terahertz}.}

Metal-dielectric metasurfaces supporting quasi-BIC resonances have been proposed for THz sensing. However, as it was shown in  \citep{Srivastava2019terahertz, Chen2020toroidal}, the  effect of losses - including both dielectric and ohmic terms - dramatically reduces the Q-factor of the resonances. 
For this reason, the use of all-dielectric structures has been proposed to overcome the limitations arising from ohmic loss. 
These metasurfaces have resonators made of moderately high permittivity materials showing low loss, such as lithium tantalate ($\rm LiTaO_{3}$) \citep{Wang2021ultrasensitive,Chen2020tunable} or silicon (Si) \citep{Tuz2018alldielectric}. In addition to the material of the resonators, metasurfaces need supporting substrates which must be chosen to have specific properties depending on the application and the fabrication technique to be used. Quartz is probably the most common material for the substrate due to being compatible with fabrication techniques in all-dielectric metasurfaces and relatively low-loss. Due to the low loss of these materials, some of these studies neglect loss when making numerical studies of their structures \citep{Wang2021ultrasensitive}. 

In this paper, we study the effect of losses in constituent materials on the performance of all-dielectric metasurfaces based on quasi-BIC and show that even small amounts of loss have a considerable impact. We also discuss the limitation of the classical figures of merit to estimate the sensing capabilities of these structures and propose an alternative definition based on our analysis.  
Finally, we compare the sensing capabilities of its high-Q modes with a periodic structure that allows extraordinary optical transmission (EOT). 

\section{Impact of material losses on the resonant properties of dielectric metasurfaces} 

\begin{figure*}
  \centering
\begin{minipage}[]{0.36\textwidth}
   \subfigure[]{
   \vspace*{-1.5in}
   \includegraphics[width=1\linewidth]{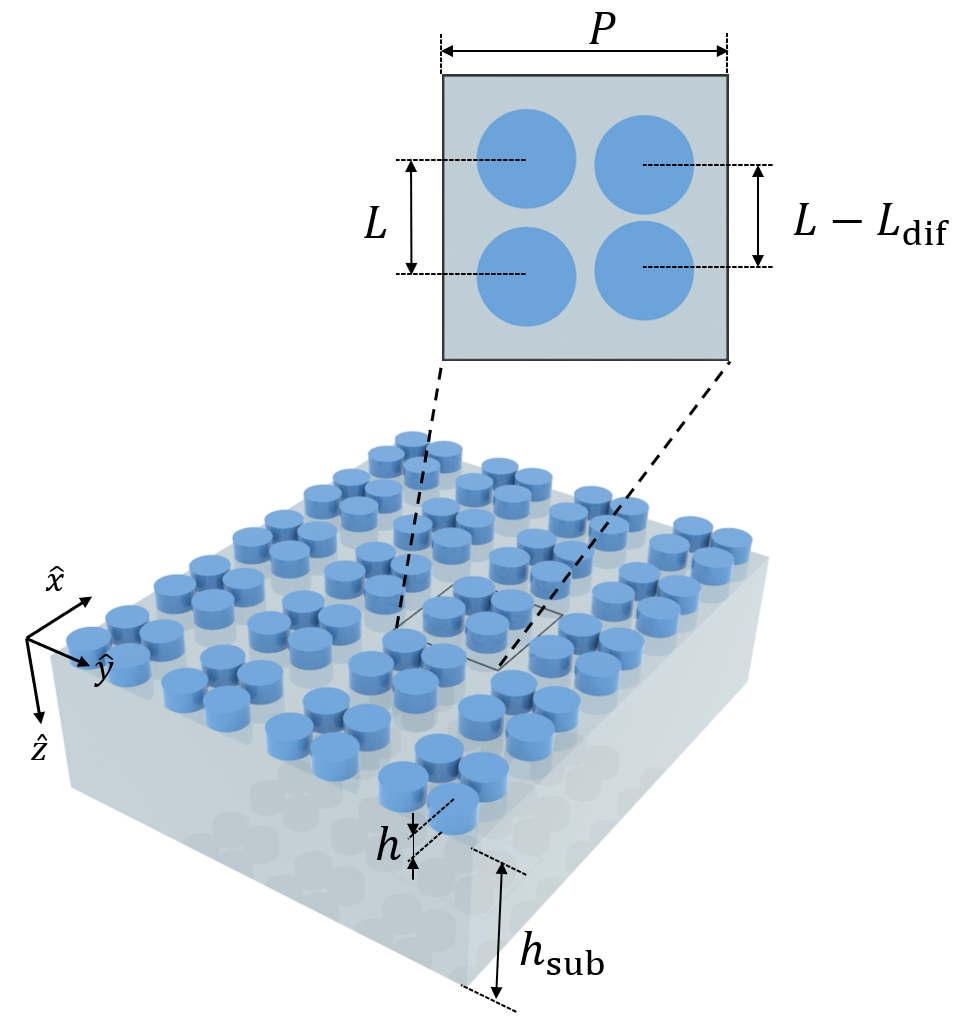}}
\end{minipage}
\begin{minipage}[]{0.6\textwidth}
     \subfigure[]{ \includegraphics[width=0.31\linewidth]{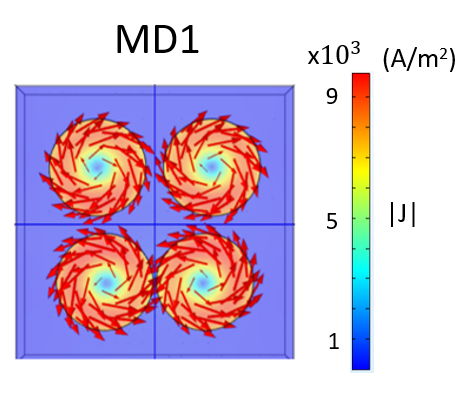}}
     \subfigure[]{ \includegraphics[width=0.31\linewidth]{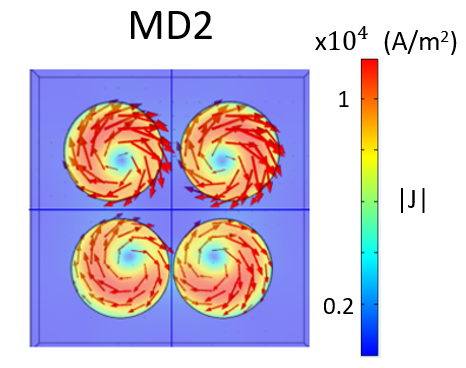}}
     \subfigure[]{ \includegraphics[width=0.31\linewidth]{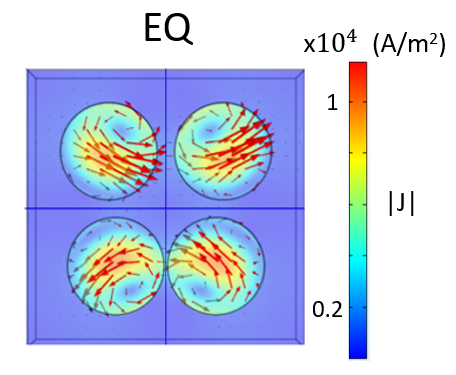}}
     \subfigure[]{ \includegraphics[width=0.48\linewidth]{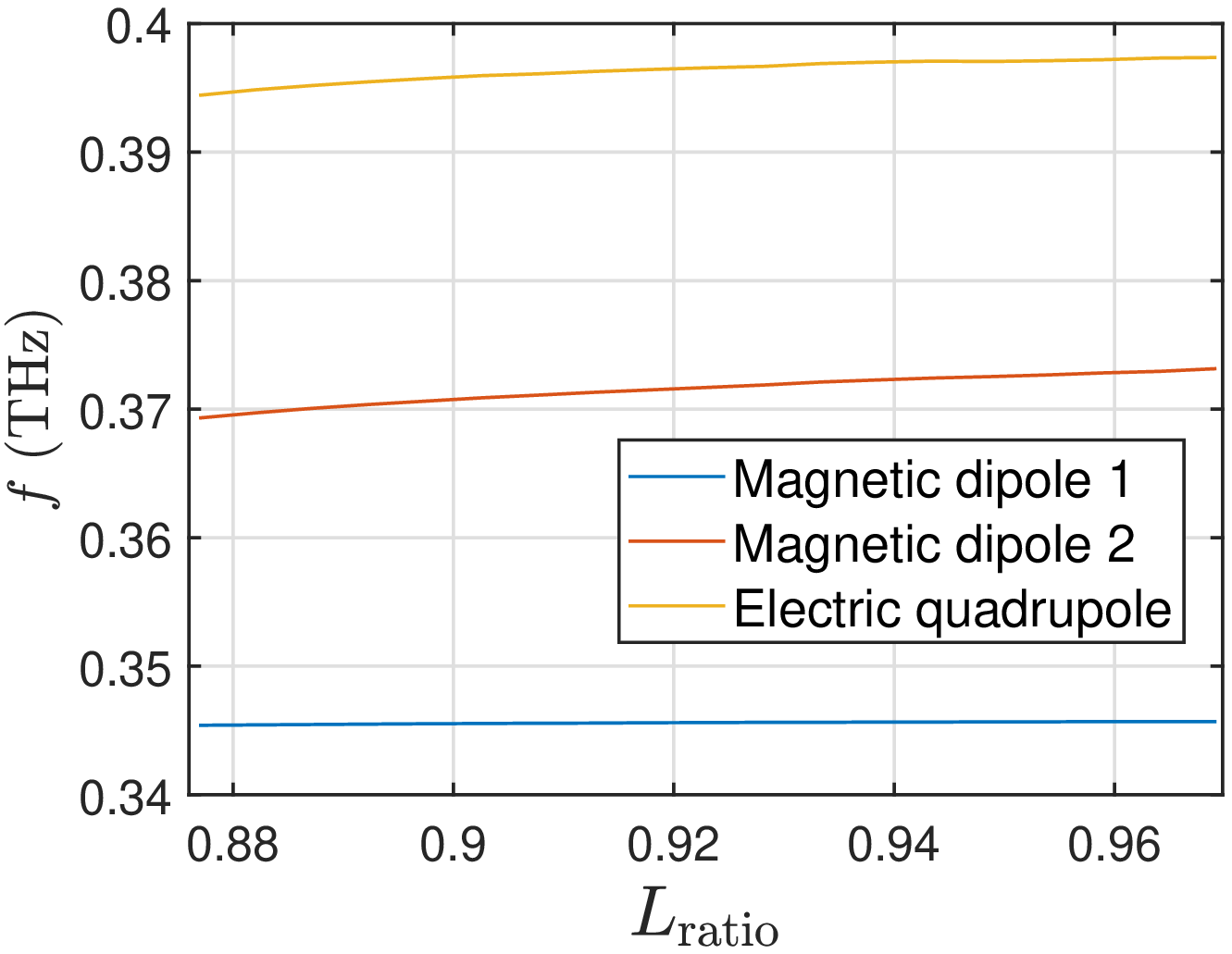}}
      \subfigure[]{ \includegraphics[width=0.48\linewidth]{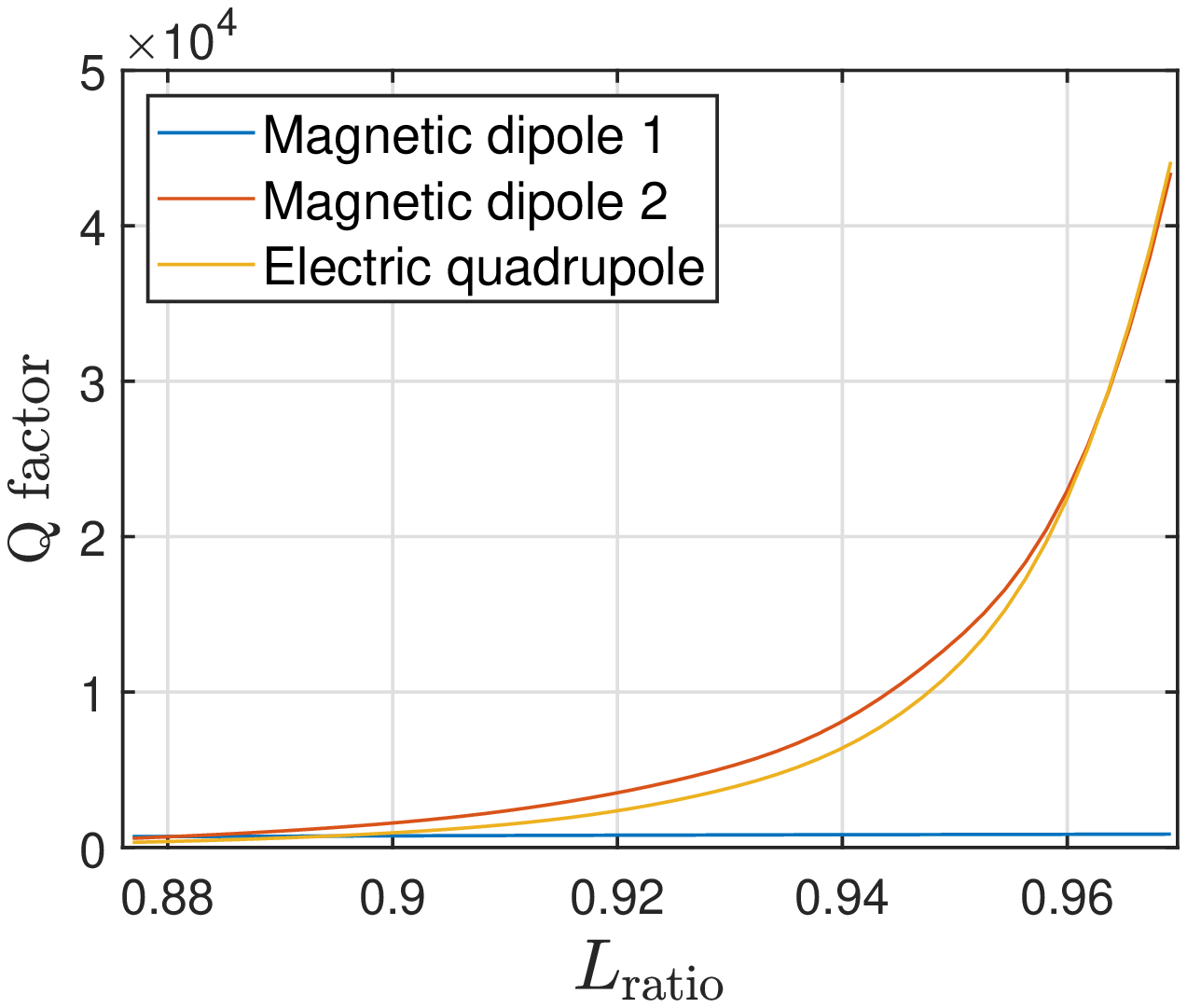}}
\end{minipage}
  \caption{\textbf{ Eigenmode study of the all-dielectric structure without losses.} (a) Schematic representation of the all-dielectric structure supporting high-Q modes. (b) Current distribution for the magnetic dipolar mode (MD1) when $L_{\rm ratio}=0.88$ and $f = 0.3454 \, \rm THz$. (c) Current distribution for the magnetic dipolar mode (MD2) when $L_{\rm ratio}=0.88$ and $f = 0.3693 \, \rm THz$. (d) Current distribution for the electric quadrupole mode (EQ) when $L_{\rm ratio}=0.88$ and $f = 0.3944 \, \rm THz$. (e) Real part of the eigen frequency as a function of $L_{\rm ratio}$.(f) Q factor for all resonant modes without losses as a function of $L_{\rm ratio}$.}
  \label{Figure1_preliminar}
\end{figure*}

In this section, we study the effect of losses in all-dielectric metasurfaces supporting strong resonant excitations in the THz band. 
To this end, we consider a metasurface based on meta-molecules consisting of four high-index dielectric cylinders [see Fig. \ref{Figure1_preliminar}(a)]. The response of such metasurfaces is underpinned by the distribution of displacement currents on the dielectric structure rather than conduction currents on metallic structures.
Each meta-molecule consists of four nanodisks, made of $\rm LiTaO_{3}$, deposited on a quartz substrate. The lattice constant is $P=378 \, \mu \rm m $ , the cylinder height is $h=66 \, \mu \rm m$, and the cylinder radius is $R=66 \, \mu \rm m$. 
The thickness of the substrate is set to $h_{\rm subs}=500 \, \mu \rm m$. 
The cylinders are arranged to form an isosceles trapezoid with a height and long side $L= 156 \, \mu \rm m$, and a short side $L_{\rm short}=L-L_{\rm dif}$, where the symmetry line is along the $y$-axis. 
The quartz substrate is considered to have $n=2$ as the real part of the refractive index of quartz, as in \citep{Wang2021ultrasensitive}, while we used $k=0.0025$ as the imaginary part of the refractive index, obtained by extrapolating the data in \citep{Davies2018temperaturedependent} to the frequency range of the resonance. 
As for $\rm LiTaO_{3}$, we use the Lorentz-type dispersion formula $\epsilon = \epsilon_{\rm \infty}(\omega^2 - {\omega_{\rm L}}^2 + i \omega \gamma) / (\omega^2 - {\omega_{\rm T}}^2 + i \omega \gamma)$, where the dielectric permittivity for high frequencies is $\epsilon_{\rm \infty}=13.4$, the longitudinal optical phonons frequency is \textcolor{black}{$\omega_{\rm L}/2\pi =7.46 \, \rm THz$, the transverse optical phonons frequency is $\omega_{\rm T}/2\pi =4.25 \, \rm THz$, and the damping factor is $\gamma/2\pi = 0.15 \, \rm THz$ \citep{Schall1999farinfrared, Wheeler2005three}.} 
This structure has been selected due to its high theoretical Q-factor and the presence of several modes to ease comparison. 
Notice that a similar structure has been recently proposed as an ultrasensitive sensor in the THz band \citep{Wang2021ultrasensitive}.

As it was demonstrated in \citep{basharin2015dielectric}, the combined EM field profile of the individual responses in the cylinders resembles a flow similar to that existing in multi-poles.
Similarly, when the meta-molecules are arranged in two-dimensional arrays, it is possible to excite high-Q multi-poles whose properties can be controlled through the asymmetry of the structure, i.e., the interaction between the nanodisks in the meta-molecules \citep{Tuz2018alldielectric}. For this reason,  a cluster of high-index dielectric nanodisks  can  provide not only electric and magnetic dipole responses but also strong magnetic quadrupole response or toroidal dipole response.  
To enhance the response of multi-poles, the dimension of the array and the cylinders are engineered to originate trapped (dark) modes for a given illumination. 
To understand the effect of losses on these resonant modes, the first step in our analysis is to identify the resonance modes supported by the structure and their resonance properties (resonant frequency and Q-factor) in the lossless scenario.

Using an eigenmode analysis of one unit cell in COMSOL and setting the dielectric loss tangent of the constitutive materials to zero, we can easily identify and characterize the resonant modes for different values of $L_{\rm ratio}=L_{\rm short}/L$.  
For $L_{\rm ratio}=1$, where the cylinders are forming a square, there is only one resonance in our region of interest. This resonance is related to a magnetic dipole (MD1) excitation in each cylinder, where two of the magnetic dipoles have the opposite orientation to the other two. This resonance is also present for $L_{\rm ratio}\neq 1$ but, in this case, two resonances appear at higher frequencies. Figure \ref{Figure1_preliminar}(b) and Figure \ref{Figure1_preliminar}(d) show the representation of these modes when $L_{\rm ratio}=0.88$. The second resonance is also related to the magnetic dipole excitation where all four dipoles follow the same orientation. It is important to notice that, due to the symmetry of the fields, this resonant mode cannot be excited by a plane wave at normal incidence when $L_{\rm ratio}=1$. The other one is a resonance related to the excitation of an electric quadrupole (EQ) as the combination of the individual excitation of dipolar moments in each cylinder. 
It is important to mention that, as it was presented in \citep{Wang2021ultrasensitive}, the structure can, for some specific conditions, support the so-called toroidal modes. 

Figure \ref{Figure1_preliminar}(e) and Figure \ref{Figure1_preliminar}(f) show the change in the resonance frequency, $\Re(f_{\rm eigen})$, and the quality factor, $Q=\Re(f_{\rm eigen})/2\Im(f_{\rm eigen})$,  of these resonances as a function of $L_{\rm ratio}$. For all the resonances, both the frequency and quality factor increase with $L_{\rm ratio}$, although it happens differently for the first magnetic dipolar resonance and the other two resonances. The resonance frequency of MD1 has a very small shift compared to the other resonances and the quality factor also has a slow increase that tends to a finite value for $L_{\rm ratio}=1$. Meanwhile, the Q factor of the other two responses tends to infinity for $L_{\rm ratio}=1$, which is the reason why they cannot be excited when the cylinders form a symmetric square structure.

\begin{figure*}
  \centering
 \begin{minipage}[t]{0.3\textwidth}
     \subfigure[]{\includegraphics[width=.9\linewidth]{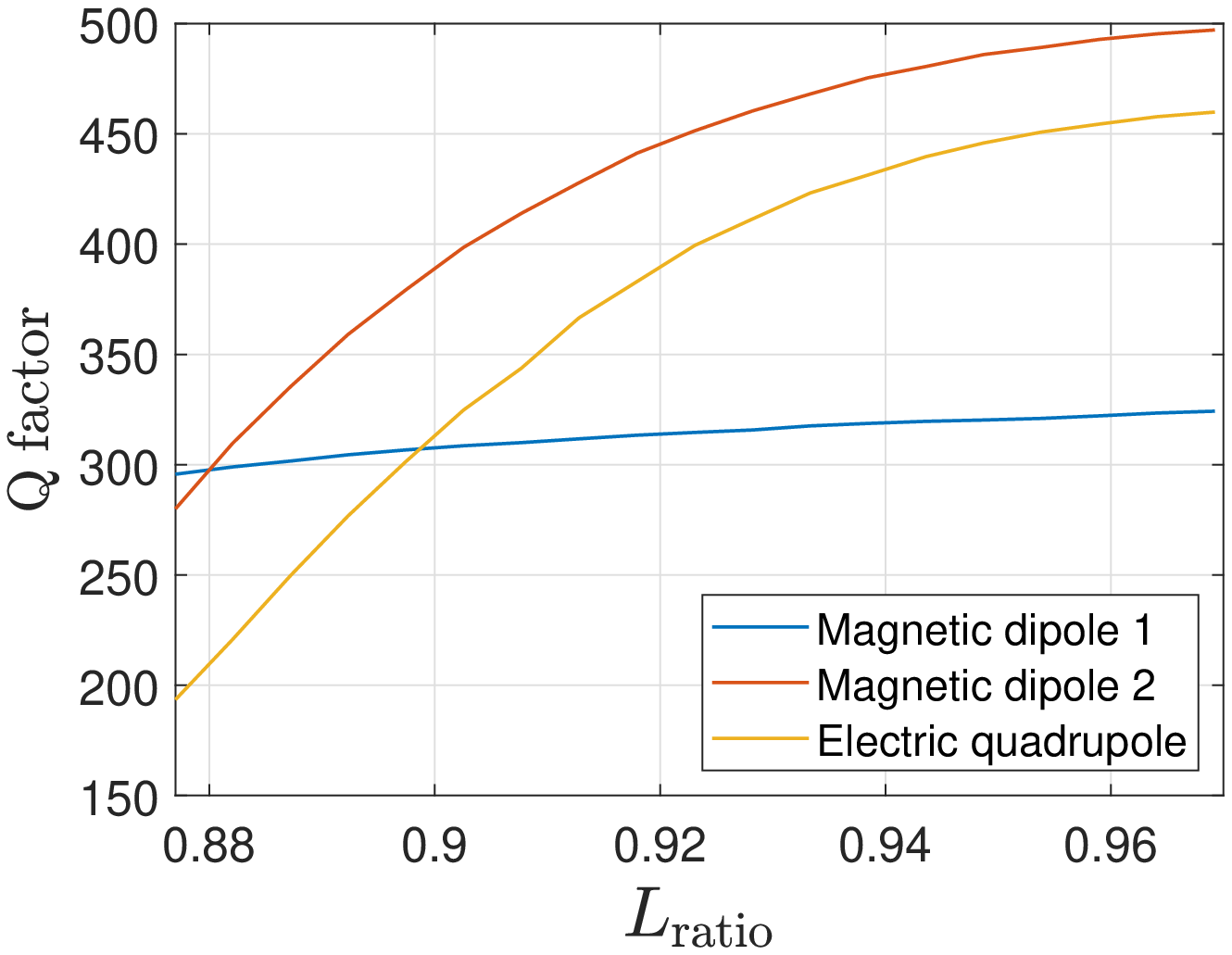} \label{Qabsall}}
    \subfigure[]{\includegraphics[width=.9\linewidth]{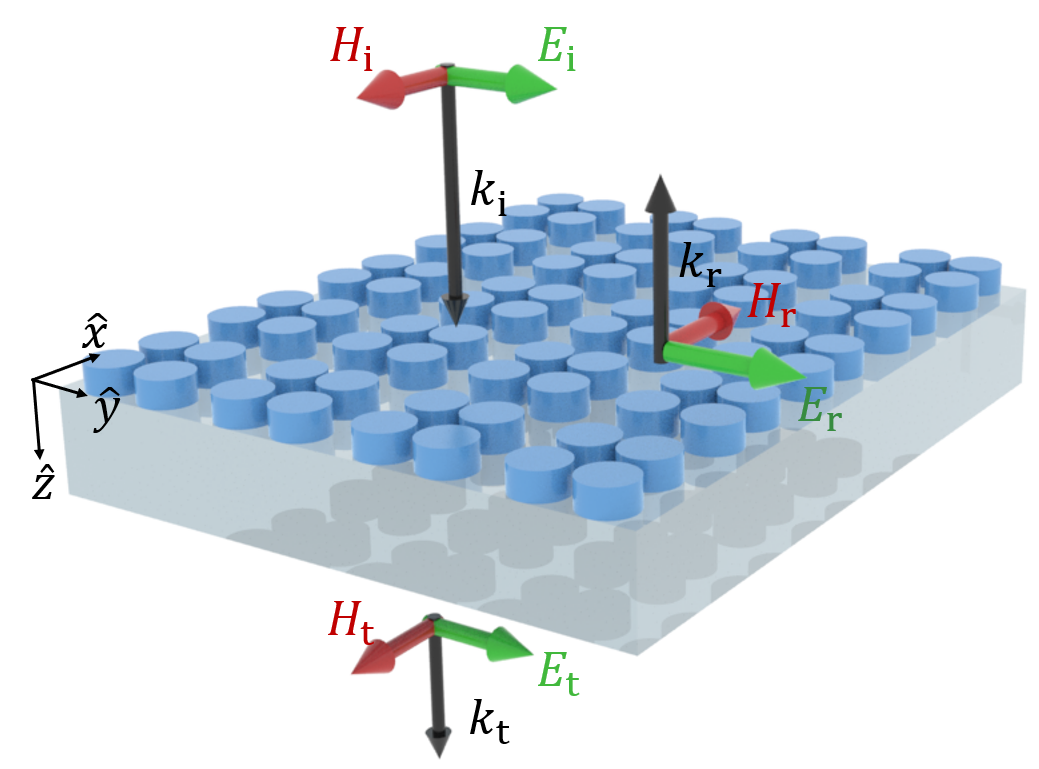} \label{Fig21}}
\end{minipage}
\begin{minipage}[t]{0.33\textwidth}
  \subfigure[]{\includegraphics[width=.8\linewidth]{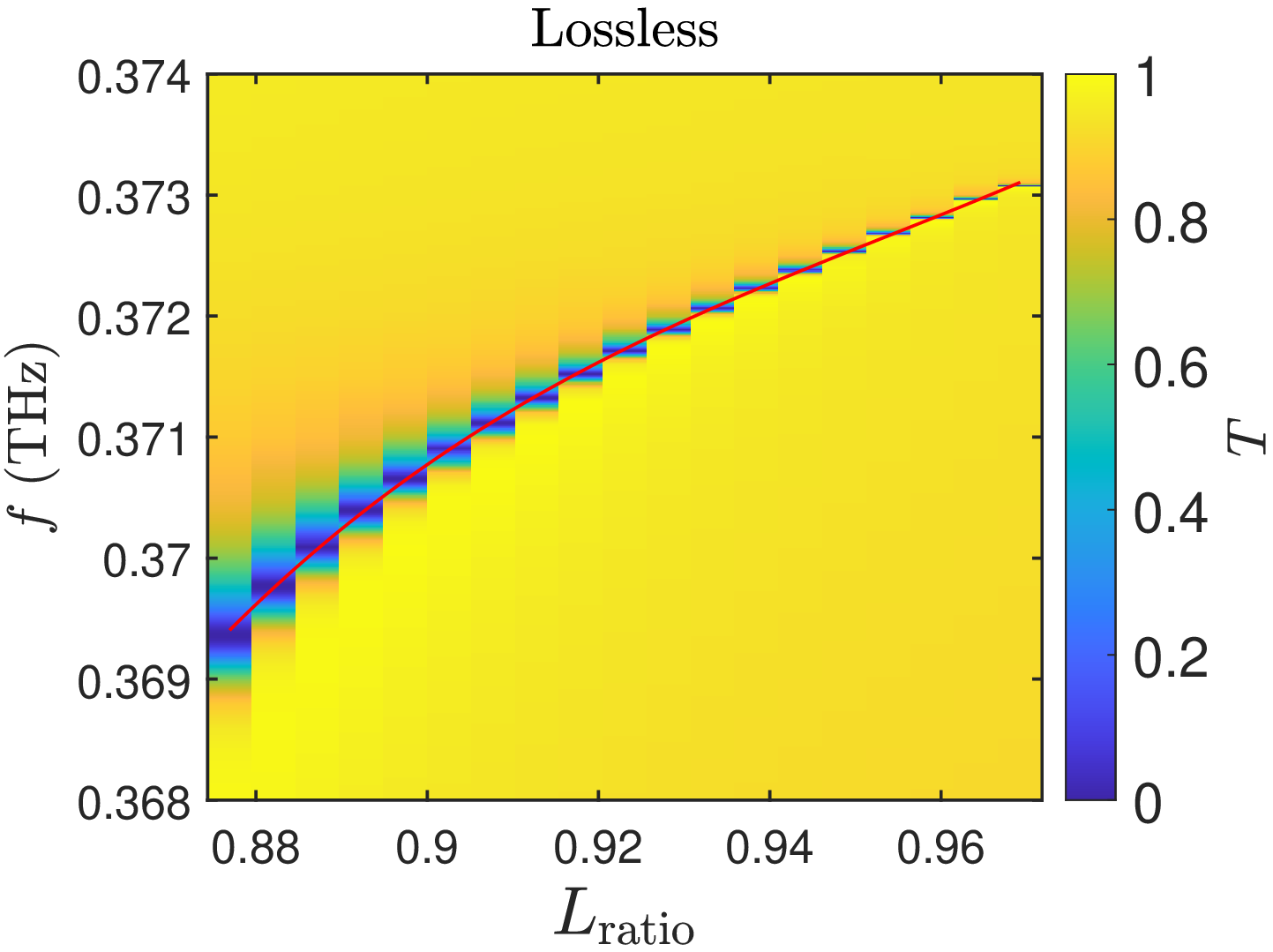} \label{Tcolormapa}}
      \subfigure[]{\includegraphics[width=.8\linewidth]{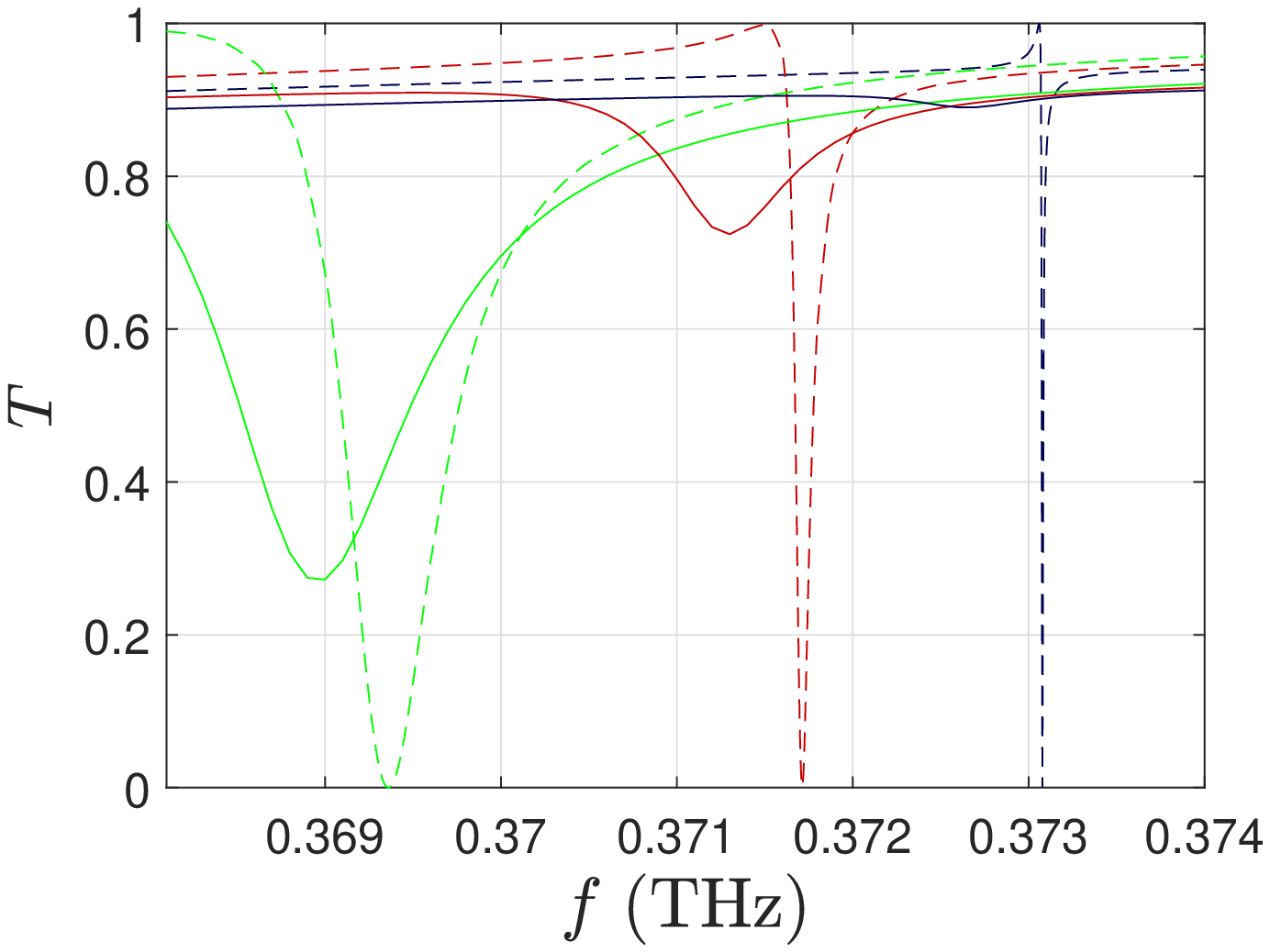} \label{RTcompa}}
\end{minipage}
 \begin{minipage}[t]{0.33\textwidth}
  \subfigure[]{\includegraphics[width=.8\linewidth]{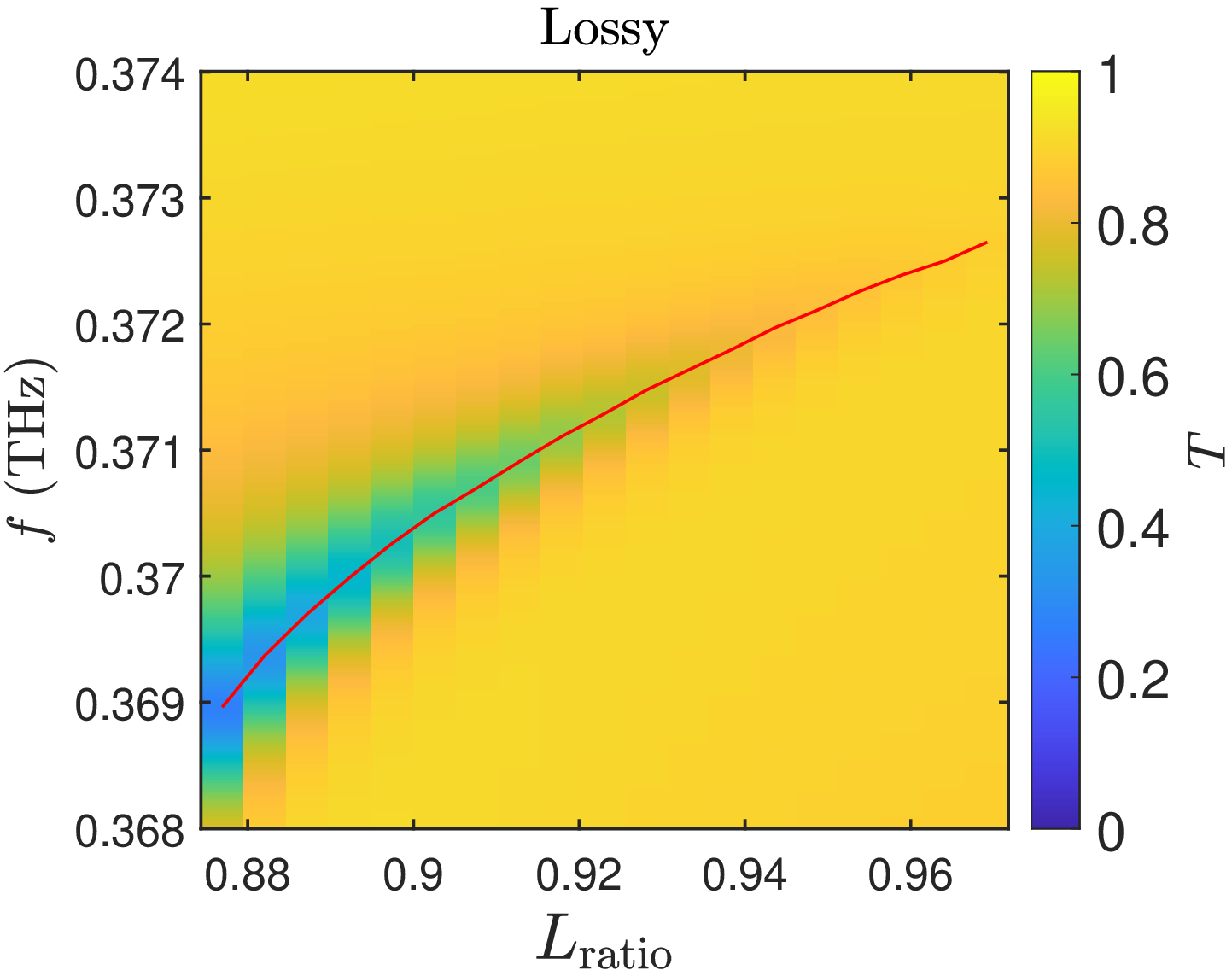} \label{Tcolormapb}}
    \subfigure[]{\includegraphics[width=.8\linewidth]{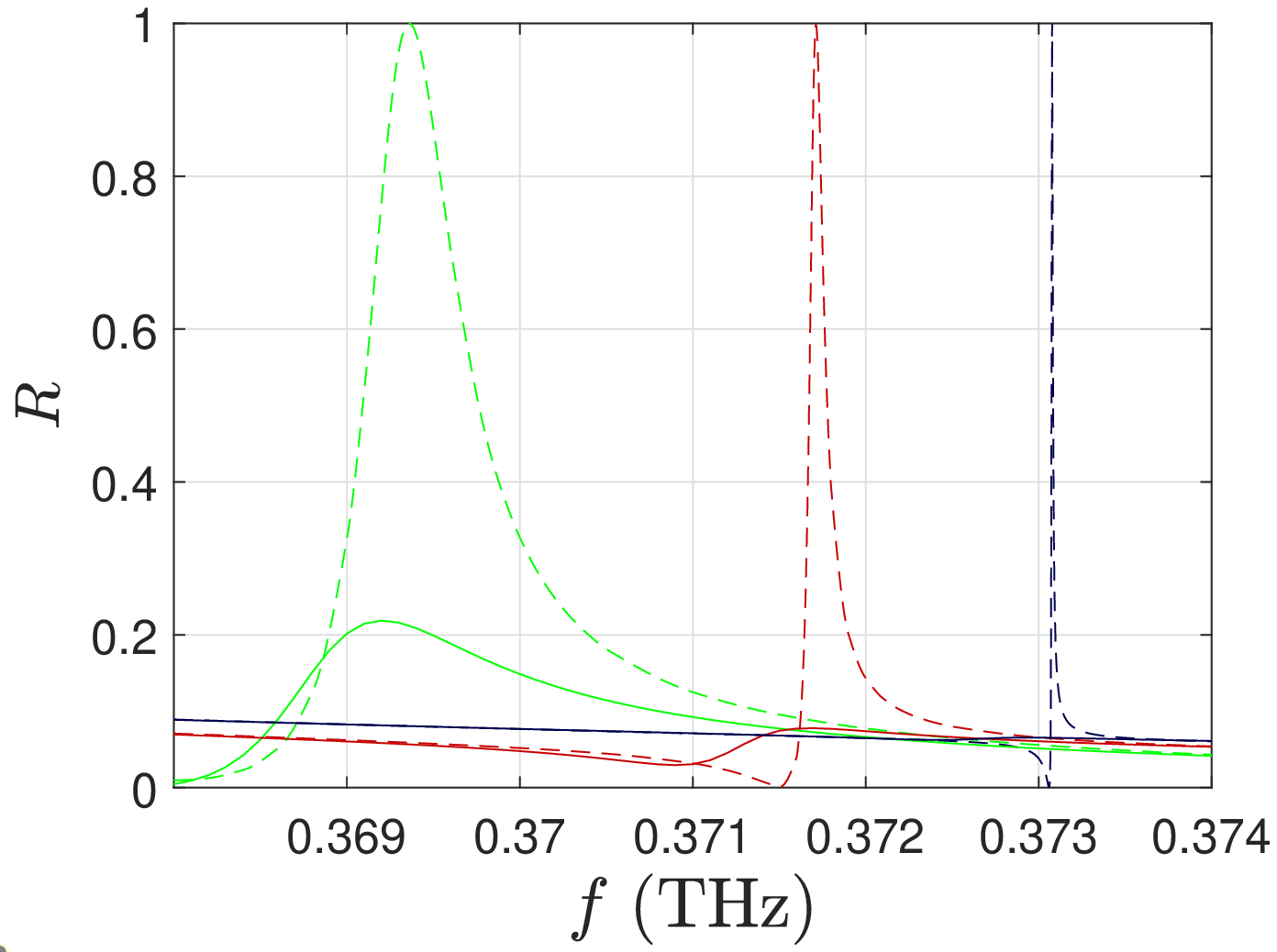}\label{RTcompb}}
\end{minipage}
  \caption{\textbf{ Study of all-dielectric modes with losses in constitutive materials.} (a) Quality factor for all resonances with losses as a function of $L_{\rm ratio}$. (b) Schematic representation of the scattering study.   (c) and (e) Colormap of transmittance as a function of frequency and $L_{\rm ratio}$ for the lossless and lossy scenario. The red curve signals the position of the resonance peak for each value of $L_{\rm ratio}$. (d) and (f) Transmission and reflection spectra for $L_{\rm ratio}=0.97$ (blue), $L_{\rm ratio}=0.92$ (red) and $L_{\rm ratio}=0.88$ (green), both for the case with losses (solid curves) and the one without (dashed curves).}
   \label{Figure2}
\end{figure*}

Once the resonant properties of the lossless structure have been characterized, the next step is to take into account losses in both the substrate and cylinders.  
We start the study by analyzing the effect of losses on the quality factor of the resonances. Figure \ref{Figure2}(a) shows the quality factor as a function of $L_{\rm ratio}$ extracted from the eigenmode analysis when losses in dielectric materials are considered. 
We can see that, aside from lowering the quality factor of all resonances, the maximum value of the quality factor tends to a finite value for all three resonances when $L_{\rm ratio}$ gets closer to unity.
Considering the losses, the resonance MD2 is the one that reaches the highest quality factor. 
As we mentioned, the sensing capability of a metasurface to detect changes in the refractive index of the substance under study is typically attributed to the capacity to support resonances with a high quality factor.
In the following steps of this research, we centered our study on MD2 resonance since it is the most interesting one for sensing applications.

After the eigenmode analysis, we studied the scattering properties under plane wave  illumination  in the normal direction as it is represented in Fig. \ref{Fig21}. In particular, for exciting the desired resonance we assume that the electric field is oriented in the $x$-direction. For this study, we evaluate the transmission and reflection spectra for different values of $L_{\rm ratio}$. A comparison between the transmission spectra for the lossless and lossy cases is represented in Fig. \ref{Tcolormapa} and Fig. \ref{Tcolormapb}, respectively. For the lossless case, the transmission is zero at the frequencies of resonance and the resonance moves to higher frequencies for larger values of $L_{\rm ratio}$. The results of this study are in agreement with the eigenmode analysis in Fig. 1(f) and show that the width of the resonances decreases for larger values of $L_{\rm ratio}$, i.e. the quality factor increases. However, we can see different behaviour in the lossy case, as it is shown in  Fig. \ref{Tcolormapb}. When considering losses in the dielectric materials, two effects appear on the transmission spectra. First,  the resonances are wider than the resonances in the lossless case [see Fig. \ref{Qabsall}], as expected from the eigenmode analysis with losses.  Second, the transmittance amplitude increases at the resonance frequency when $L_{\rm ratio}$ increases or, in other words, the resonance produces smaller variations in the transmission spectra. 

\begin{figure*}
  \centering
\subfigure[]{\includegraphics[width=.35\linewidth]{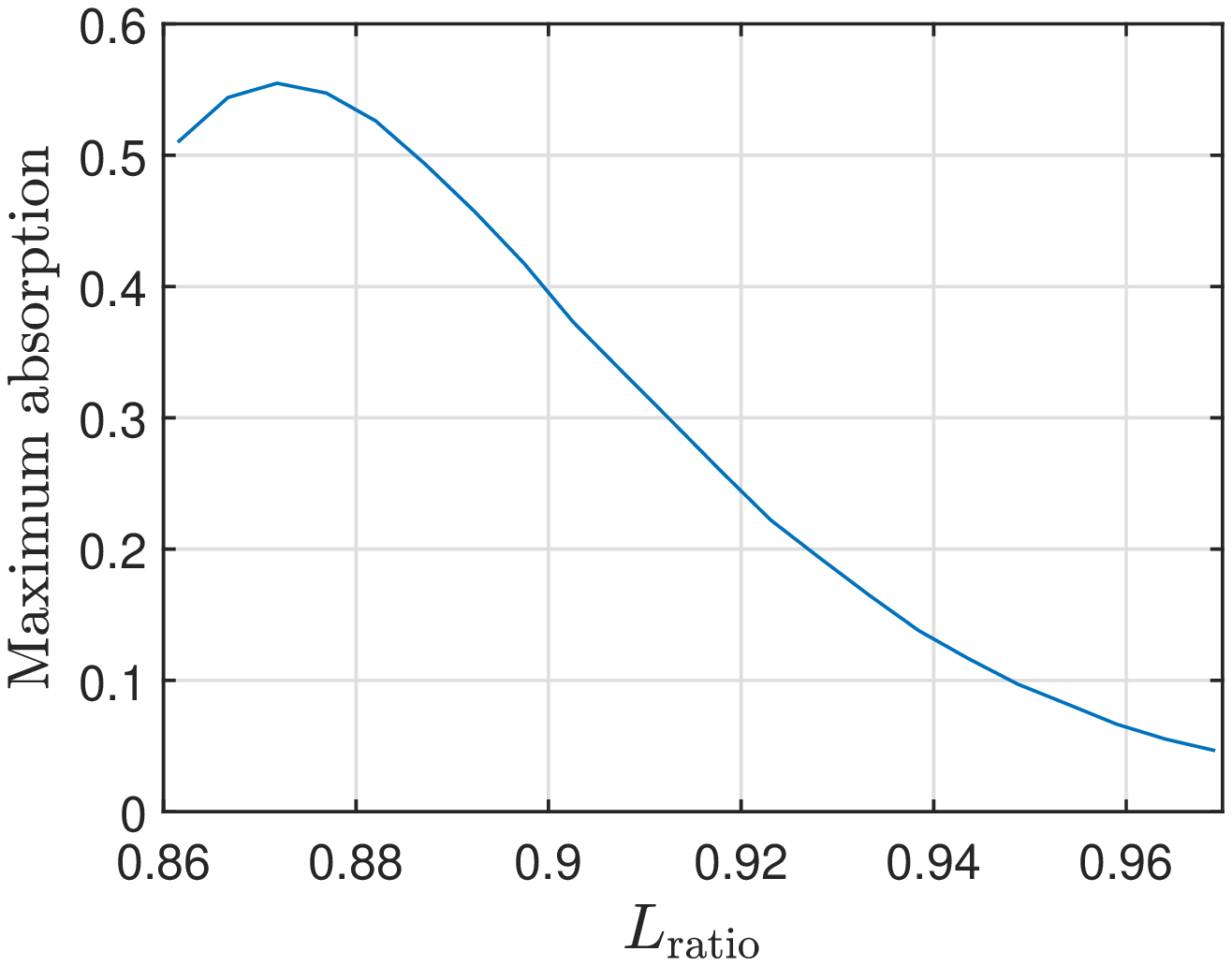}\label{fig:Abslratioa}}
\subfigure[]{\includegraphics[width=.15\linewidth]{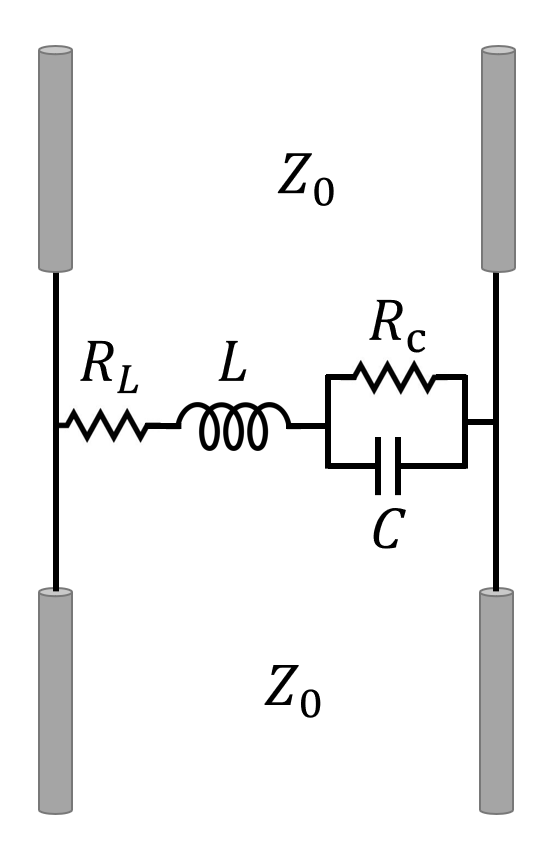}\label{fig:Abslratiob}}
\subfigure[]{\includegraphics[width=.35\linewidth]{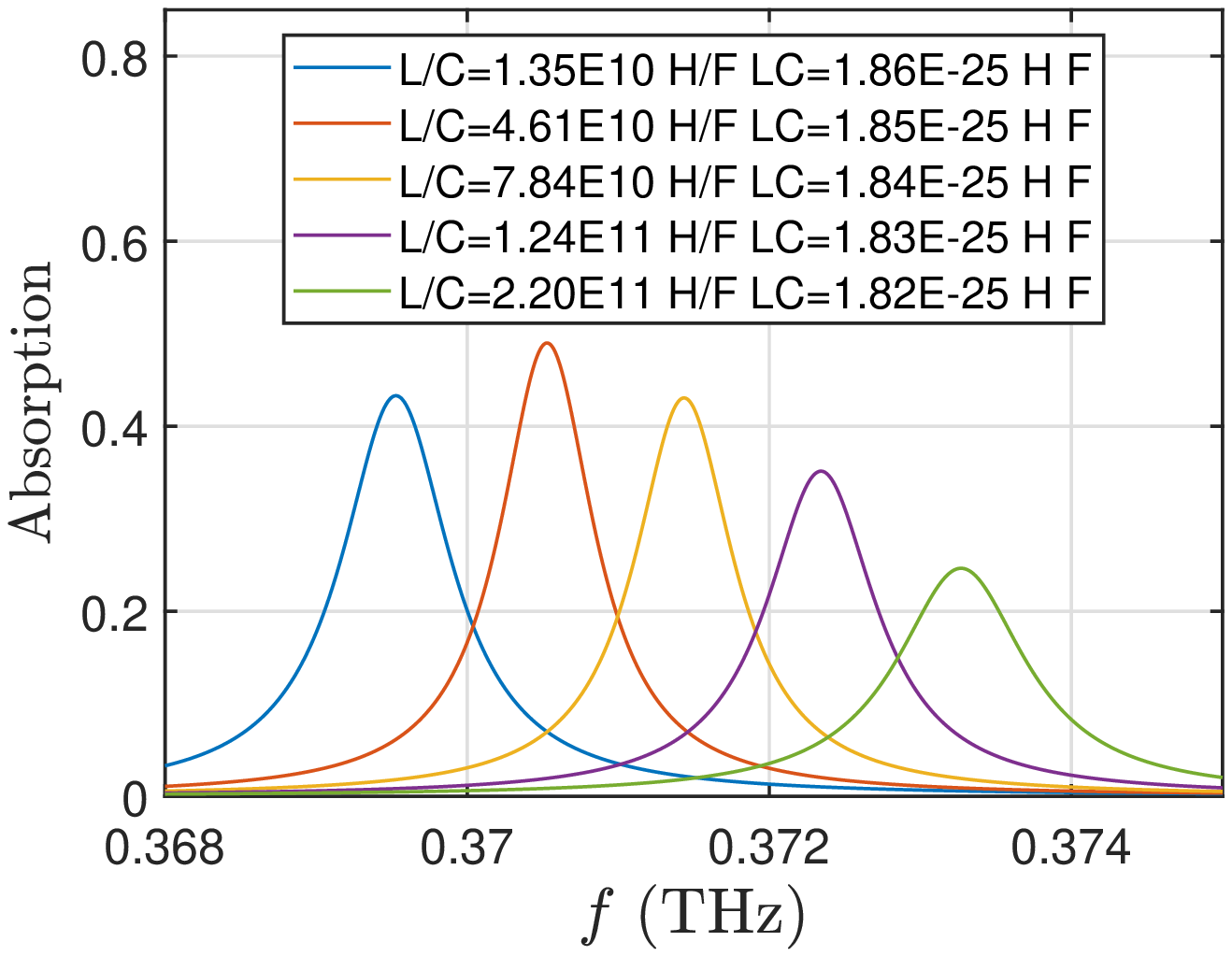}\label{fig:Abslratioc}}
\caption{\textbf{ Study of absorption.} (a) Maximun value of the absorption in the dielectric metasurface for different values of $L_{\rm ratio}$. (b) Diagram of lossy $LC$ resonant circuit. (c) Absorption resonances of the lossy LC circuit for different values of L and C, with $R_C=2\cdot 10^8 \, \Omega$ and $R_L=20 \, \Omega$. }
\label{fig:Abslratio}
\end{figure*} 

For a deeper analysis of this effect, we also evaluate the effect of $L_{\rm ratio}$ on the reflection spectra. 
Figure \ref{RTcompa} and Figure \ref{RTcompb} show the transmission and reflection spectra  for the ideal lossless structure and the structure with losses  when $L_{\rm ratio}=$ 0.97, 0.92, and  0.88. When the structure is more asymmetric ($L_{\rm ratio}=0.88$), \textcolor{black}{the maximum transmission goes from 0 in the lossless case to around 0.3 in the lossy case}, while the maximum reflection changes from unity to \textcolor{black}{0.2} due to absorption in the structure.

For larger values of $L_{\rm ratio}$, the variation of both transmission and reflection \textcolor{black}{spectra are much larger}, being the reflection spectrum more sensitive to the losses in the dielectric materials.  
From this analysis, it is clear that the variations in the transmission and reflection spectra depend on $L_{\rm ratio}$, to the point that the resonance is barely visible for larger values of $L_{\rm ratio}$ that correspond to the highest values of $Q$. This imposes a limit on how small the value of $L_{\rm ratio}$ can become, and therefore how high the value of $Q$ can be before the structure stops being suitable for sensing.

To complete the analysis, we explore the effect of losses on absorption at the resonant frequency for different configurations of the metasurface. 
Figure \ref{fig:Abslratioa} shows the maximum value of absorption for different values of $L_{\rm ratio}$. We can see that, despite the increase of the quality factor of the resonance when $L_{\rm ratio}$ gets closer to unity, the absorption has a maximum for a certain value of $L_{\rm ratio}$.
 This behaviour is common for resonant systems with lossy elements. For example, one can consider a  simple LC resonant circuit where both the $L$ and $C$ elements are considered to be real components with losses [see Fig. \ref{fig:Abslratiob}]. We can assume that a series LC-resonator is a shunt connection in a transmission line that can be modelled by the impedance $Z_t=R_L+j\omega L+\frac{R_C}{1+j\omega C R_C}$ with $R_L$ and $R_C$ measuring the losses in the inductor and the capacitor.
 The scattering parameters of this two-port network can be easily calculated as \citep{Pozar}
 \begin{subequations}
\begin{equation}
S_{21}=\frac{-2 Z_0 Z_t }{2 Z_0 Z_t + Z_0^2}
\end{equation}
\begin{equation}
S_{11}=\frac{-Z_0^2}{2 Z_0 Z_t + Z_0^2}    
\end{equation}
\end{subequations}
where $S_{21}$ and  $S_{11}$ represent the transmission and reflection coefficients. 
Finally, the expression of the absorption reads  
\begin{equation}
A=1-|S_{21}|^2-|S_{11}|^2=1-\frac{4 Z_0^3 \Re(Z_t) }{|2 Z_0 Z_t + Z_0^2|^2}    
\end{equation}

Considering low-loss scenario, i.e., when $R_L$ is very small and $R_C$ is very large,  the resonance frequency of the system can be approximated as $f_0=\frac{1}{2\pi\sqrt{LC}}$ and the quality factor will be proportional to $\sqrt{\frac{L}{C}}$. 
Let us assume constant values for $R_C=2\cdot10^8 \, \Omega$ and $R_L=20 \, \Omega$. 
 The values of $L$ and $C$ are chosen to make the resonance move slightly to higher frequencies as it happens in the metasurface under study by making  the product $L C$ decrease. At the same time, by making the ratio $\frac{L}{C}$ bigger we ensure that the resonance would have a larger quality factor than for higher frequencies. In Figure \ref{fig:Abslratioc}, we show the absorption near the LC circuit resonance for different values of $L$ and $C$.
 As it happens in our metasurface,  we see that there is a value of $\frac{L}{C}$ for which the absorption peak reaches its highest value. At this point the current in the resonator is at maximum, indicating strong interaction with the resonator. Coming back to the metasurface studied in this work and applying the same reasoning,  
  the absorption value can estimate the strength of the light-matter interaction.
From this study, we can conclude that attempting to increase the quality factor of a resonance in a system with losses can lead to weaker light-matter interaction that could result in less sensitivity.

\begin{figure}
  \centering
    \vspace*{-0.1in}
 \subfigure[]{\includegraphics[width=.6\linewidth]{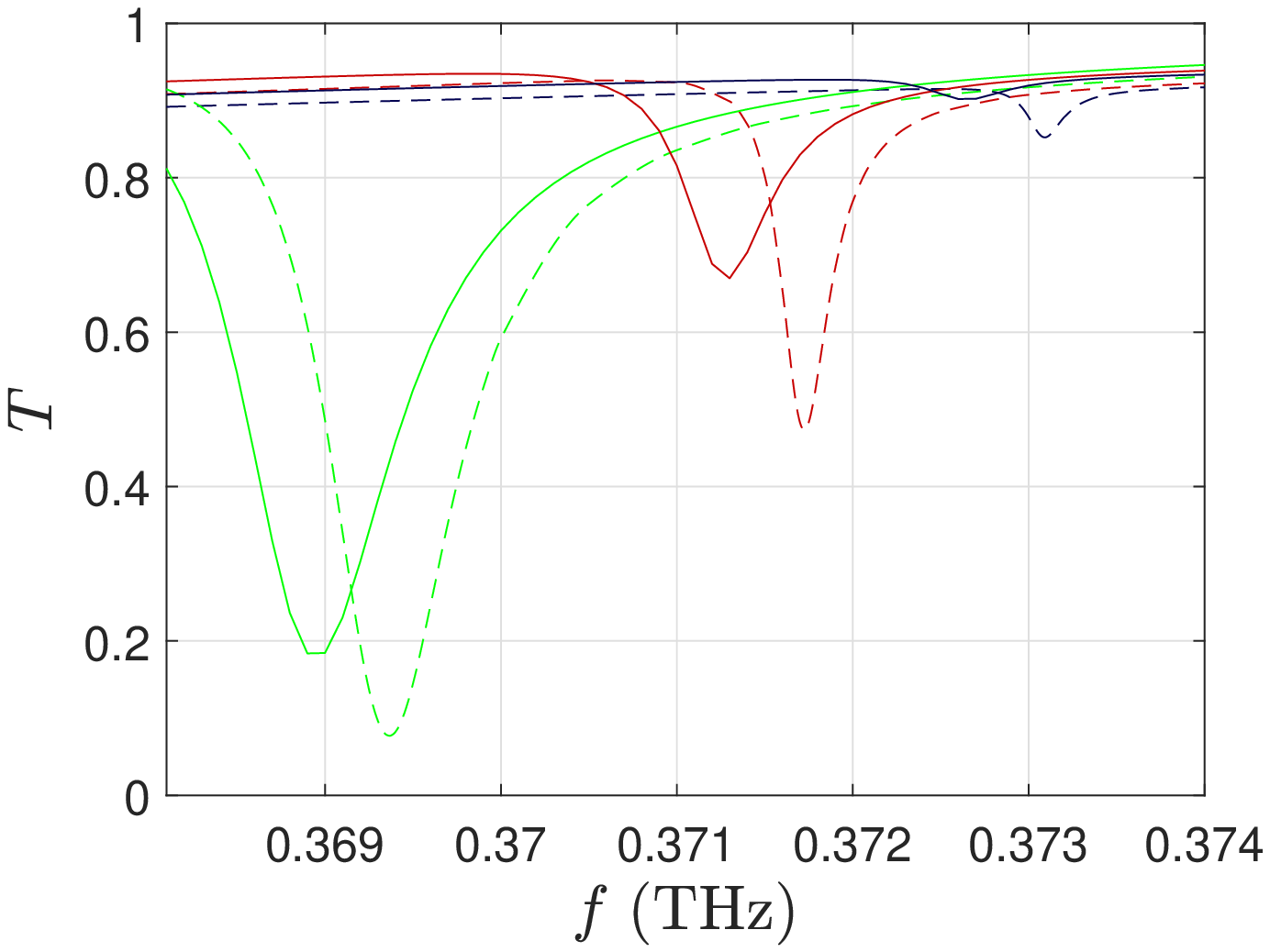} \label{fig:Subscompa}}
  \vspace*{-0.1in}
   \subfigure[]{\includegraphics[width=.6\linewidth]{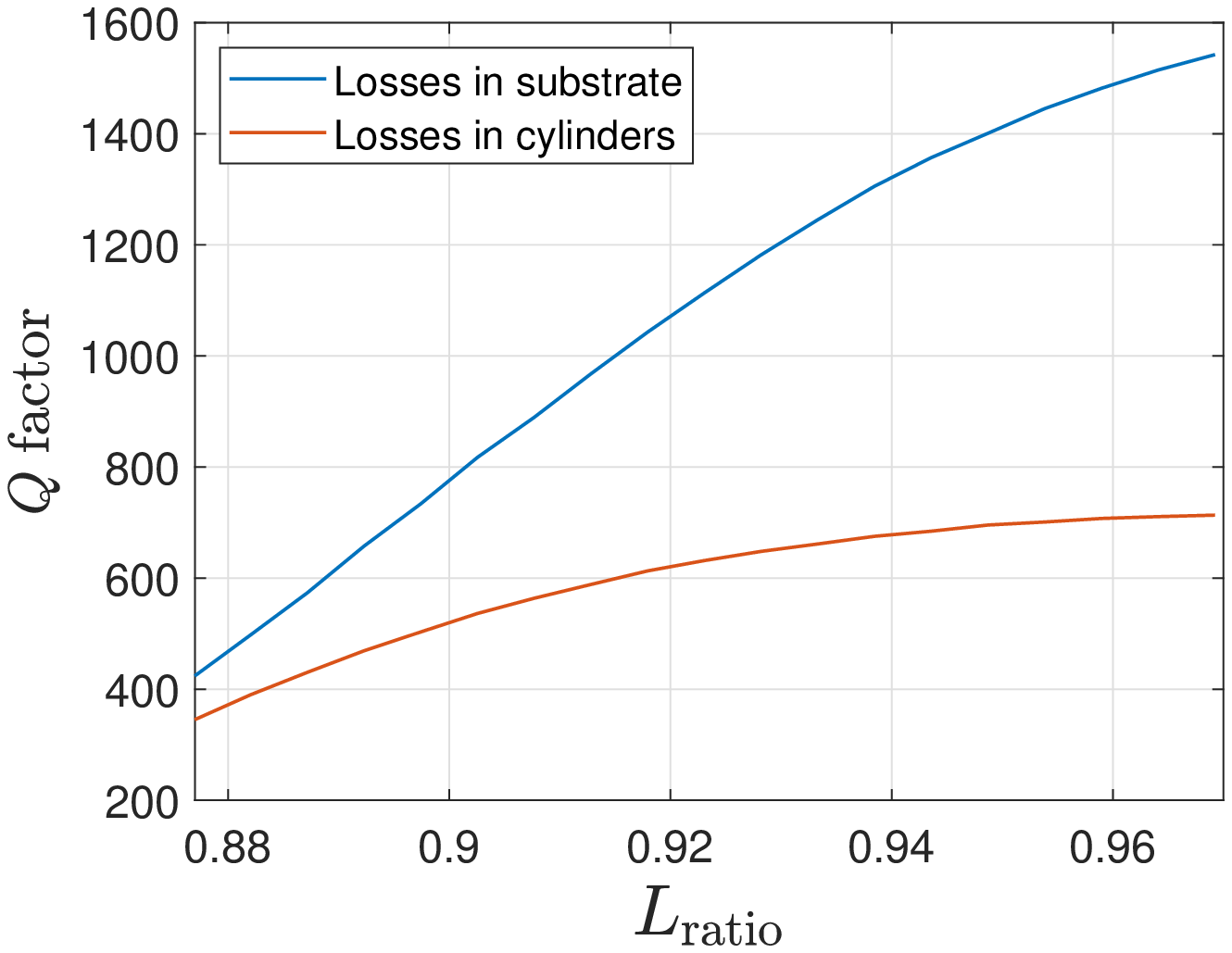}\label{Qnoabsall}}
\caption{\textbf{ Study of losses in the substrate and the cylinders.} (a) Frequency spectra of the system with absorption only in the substrate (dashed curves) and only absorption in the cylinders (solid curves) for $L_{\rm ratio}=0.97$ (blue), $L_{\rm ratio}=0.92$ (red) and $L_{\rm ratio}=0.88$ (green). (b) Q as a function of $L_{\rm ratio}$.}
   \label{fig:Subscomp}
\end{figure}

Once the analysis in this realistic scenario with losses in all constituent materials is done, it is important to analyze in more detail the effect of losses in each material individually. By doing that we will have information on how to minimize the effect of losses or and what is the optimal regime of using these in practical applications.
Figure \ref{fig:Subscompa} compares the resonance at different values of $L_{\rm ratio}$ for the structure with losses only in the cylinders and with losses only in the substrate.
Solid lines represent the transmission spectra for three different values of $L_{\rm ratio}$ when the substrate is lossless and the cylinders are lossy, \textcolor{black}{while dashed lines are used for the case with lossless cylinders and lossy substrate. We can see that the minimum values of $T$ reached follow the same trend with $L_{\rm ratio}$ as the case with losses in both elements of the system, although they are lower than in that case, as we expected since the overall losses are lower. We can also conclude that the losses in the cylinders have a higher impact on the resonance than the losses in the substrate}.

The same behavior can be seen in the study of the quality factor as a function of $L_{\rm ratio}$. Using the results from the eigenmode analysis considering losses in the materials, Fig. \ref{Qnoabsall} represents the quality factor for these two scenarios.
\textcolor{black}{While both cases reach lower values than the lossless case, the lossy substrate case shows a higher Q factor than the lossy cylinders case.}

\begin{figure*}
  \centering
    \subfigure[]{\includegraphics[width=.3\linewidth]{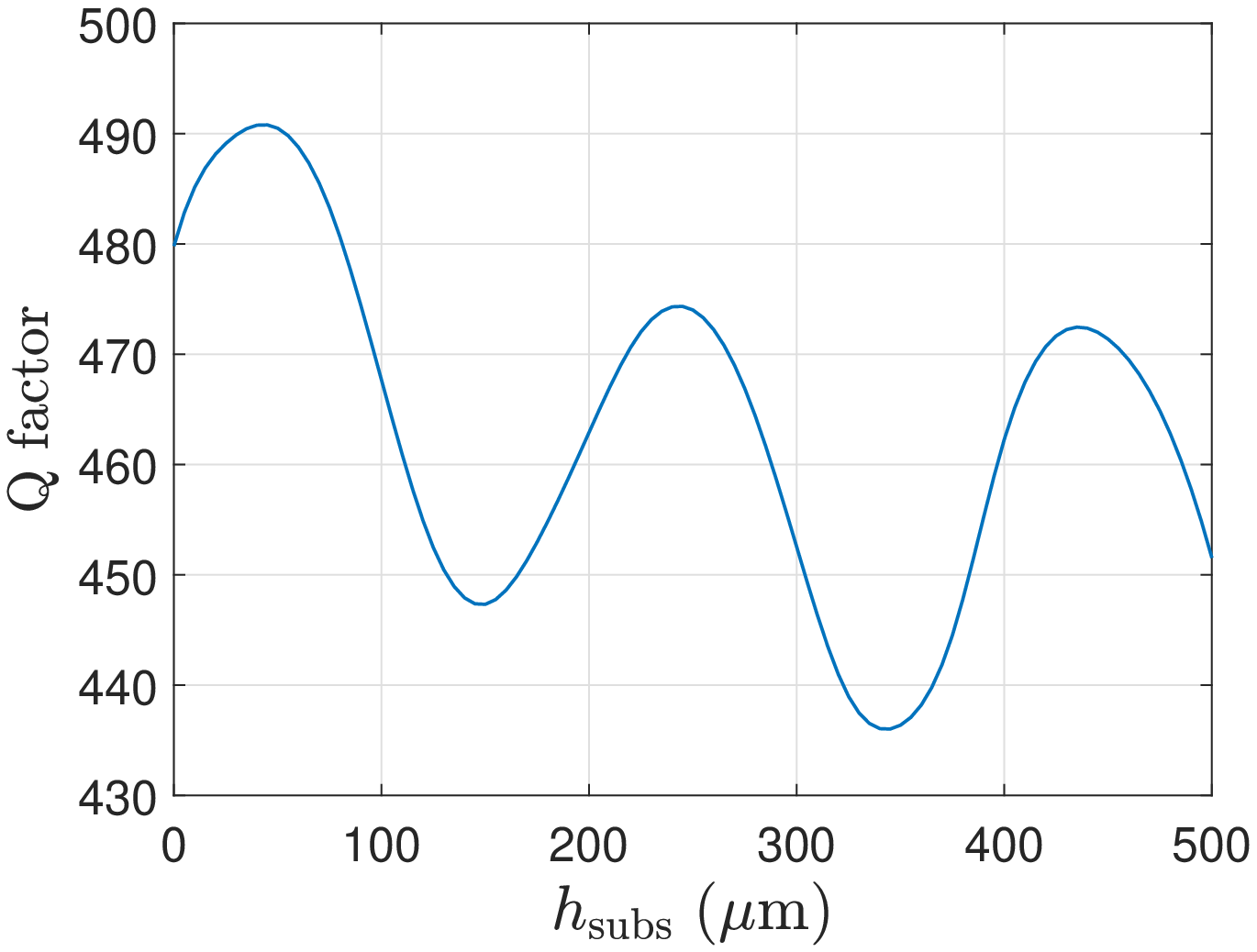} \label{fig:Qhsubsa}}
   \subfigure[]{\includegraphics[width=.3\linewidth]{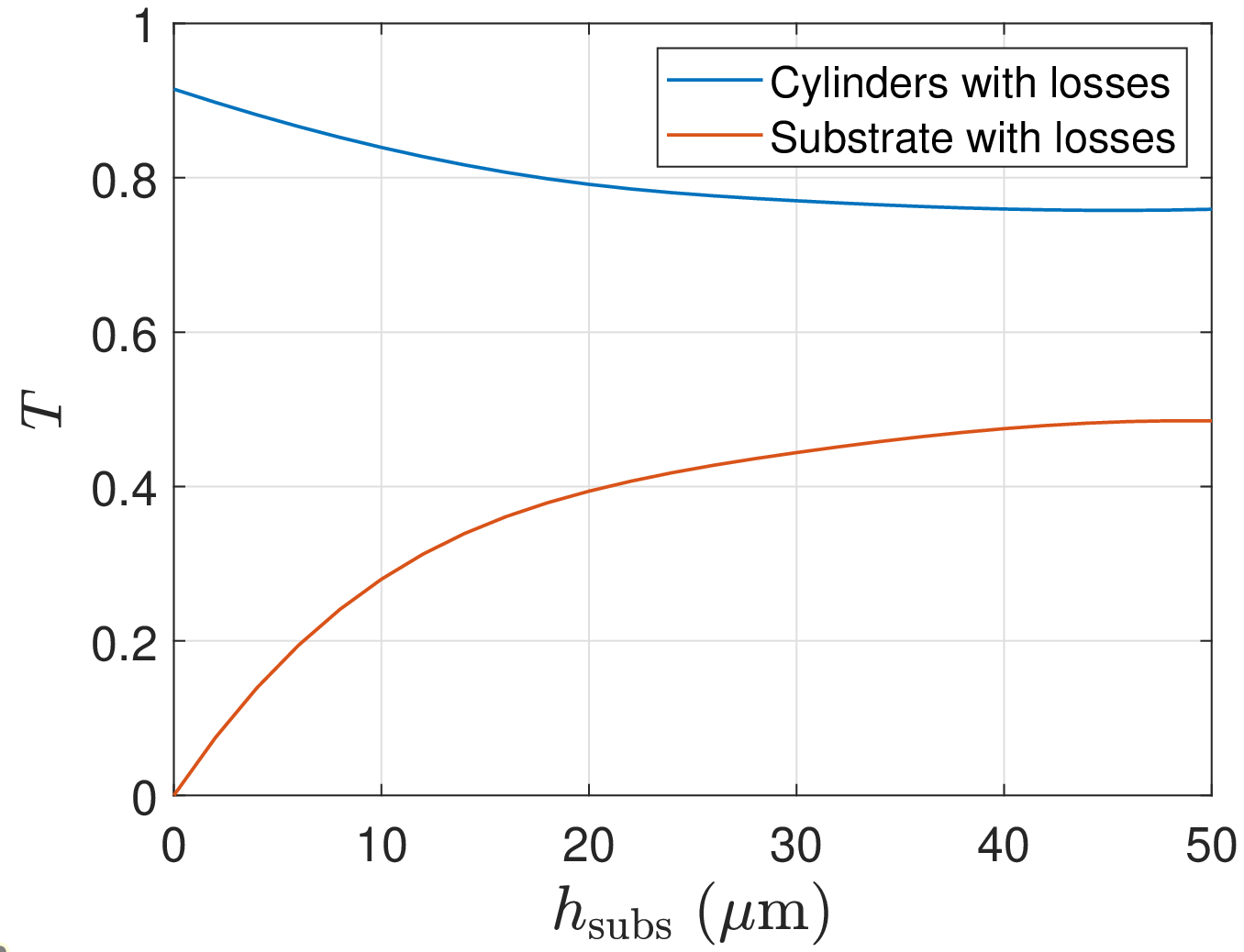}\label{fig:Qhsubsb}}
   \subfigure[]{\includegraphics[width=.3\linewidth]{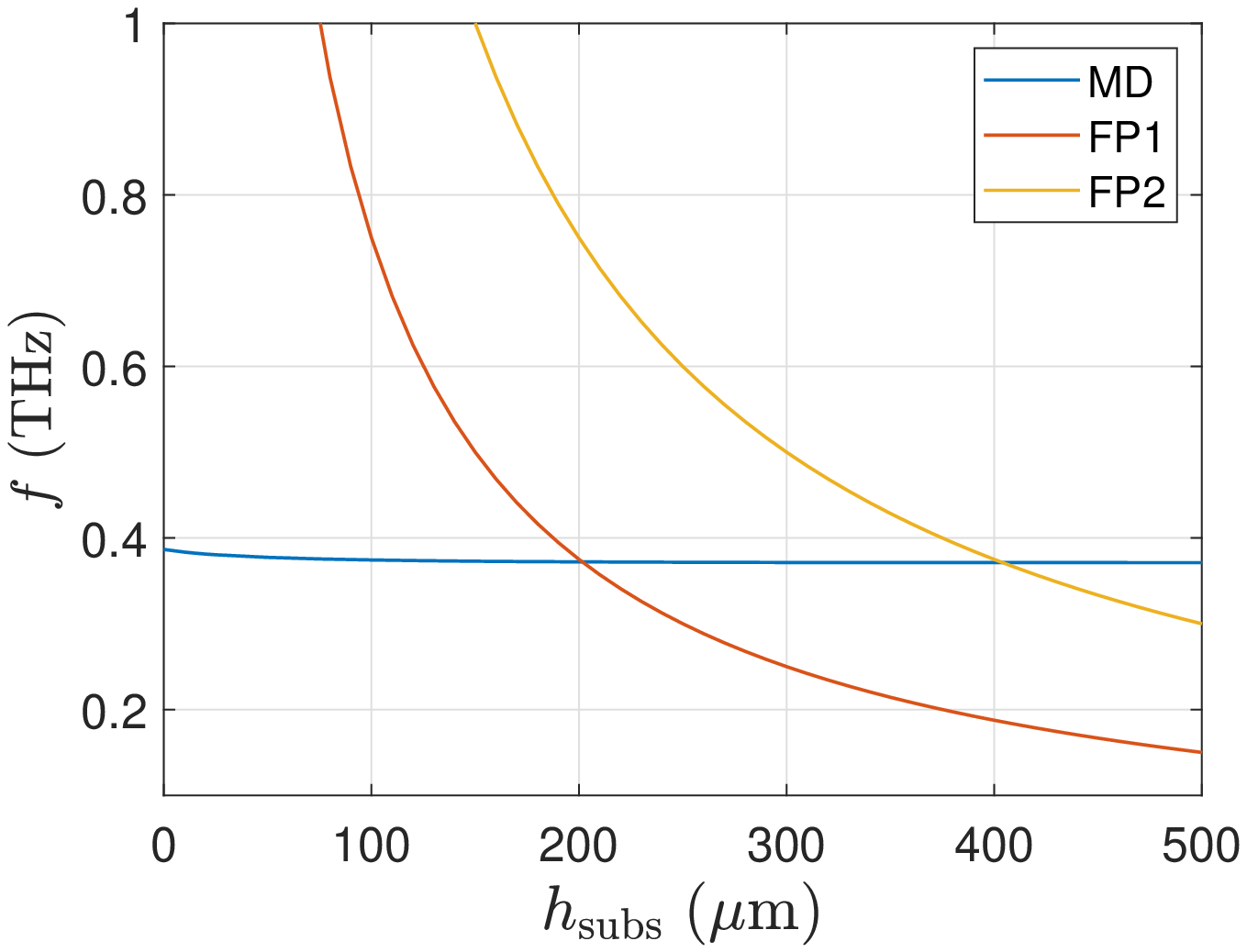} \label{fig:Qhsubsc}}
\caption{\textbf{ Study of losses for different values of substrate thickness.} (a) Q as a function of the substrate thickness $h_{\rm subs}$ for $L_{\rm ratio}=0.92$. (b) Transmission minimum for the cases of only losses in the substrate and only losses in the cylinders as a function of $h_{\rm subs}$ for $L_{\rm ratio}=0.92$ (c) Frequency position of the symmetric MD resonance and the first 2 Fabry-P\'erot resonances as a function of $h_{\rm subs}$}
   \label{fig:Qhsubs}
\end{figure*}

This result suggests that \textcolor{black}{the best path to obtain resonances with higher quality factors would be reducing the losses in the resonators. However, the materials used in this structure are already low-loss, and changing the geometry of the resonators would completely change the resonance. Meanwhile, it would be possible to reduce the losses in the substrate by} developing free-standing metasurfaces, such as the one in \citep{Fan2019dynamic}, or by using a thin membrane as a substrate, as in \citep{Peralta2009large}. \textcolor{black}{This should lead to slightly improved resonances.}
To check if this assumption was correct, we performed an eigenmode study of the structure with $L_{\rm ratio}=0.92$ for different values of the substrate thickness. Figure \ref{fig:Qhsubsa} shows the Q values obtained as a function of the substrate thickness. It can be seen that \textcolor{black}{the Q factor oscillates as a function of the substrate thickness and that the maximum value of Q factor is reached for a much thinner substrate that the one we studied previously, which we had chosen since it was a standard thickness that guarantees mechanical stability.} Despite this effect, Figure \ref{fig:Qhsubsb} shows that, while the minimum value of $T$ obtained in simulations with losses only in the substrate gets closer to 0 as the substrate thickness decreases, for the case with losses in the resonators, decreasing the substrate thickness increases the minimum value of $T$, reaching the point of having a higher value of $T$ for the case without substrate than for $h_{\rm subs}=500 \, \mu \rm m$. The reason for this property is that the field concentration in the resonators is higher in the free-standing cylinders than in the structure with a substrate due to the higher refractive index difference between the resonators and the environment, leading to a higher Q value, but also making the resonance more susceptible to losses in the resonators. \textcolor{black}{As for the oscillations in Figure \ref{fig:Qhsubsa}, they can be explained} by taking into account the Fabry-P\'erot resonances in the substrate. The equation for the transmittance through a dielectric slab of thickness \textit{d} and refractive index \textit{n} is \citep{Hecht}:
\begin{equation}
    T=\frac{1}{1+F\sin^2{\frac{\delta}{2}}}
\end{equation}
where $\delta = \frac{4 \pi}{\lambda_0} n d \cos{\theta} $, $\theta$ is the incidence angle, $\lambda_0$ is the wavelength in vacuum, and $F$ is a coefficient that depends on the reflectance of the air-dielectric interface. The transmission maxima are located in $\delta = 2 \pi m$, where $m$ is the order of resonance. In Fig. \ref{fig:Qhsubsc}, the frequency position of the resonances as a function of the substrate thickness is plotted for both Fabry-P\'erot resonances (FP) and the symmetric MD2 resonance. The values of $h_{\rm subs}$ where the FP resonances and the MD resonance coincide, while not being the same as the values for the oscillation maxima in Fig.~\ref{fig:Qhsubsa}, have the same distance between them as the distance between the oscillation maxima, making the FP resonances a plausible reason behind this phenomenon.

\textcolor{black}{Another detail that we can observe in Fig. \ref{fig:Qhsubsc} is a small change in the frequency of the resonance for low values of $h_{\rm subs}$, meaning that the resonance is affected by changes in the substrate refractive index, since making the substrate infinitely thin is equivalent to changing its refractive index from 2 ($n$ of quartz) to 1 ($n$ of air). This effect, however, is much less significant than changing the refractive index of the sample.}


\section{Performance of lossy metasurfaces as refraction index sensors} 
The sensing capability of a resonant metasurface to detect changes in the refractive index of the substance under study is typically quantified by a figure of merit (FOM). In the literature, different FOM definitions can be found depending on the scenario and the application conditions of the sensor. 
Some authors consider, at a fixed frequency, the transmission change induced by a change in the refractive index \citep{Alipour2018High,Dolatabady2017Tunable}, ${\rm FOM}=\Delta T/(T \Delta n)$ where $\Delta T/T$ is the normalized change in the transmission spectra and $\Delta n$ is the change in the refraction index. This definition is convenient for single-frequency characterization systems, such as when lasers  are used for illumination with one wavelength output.   

However, for sensing systems at THz frequencies where instruments make available broad spectra for characterization, it is more convenient to define the figure of merit as  the  ratio  between  the  sensitivity ($S=\Delta f_{\rm res}/\Delta n$)  and  the  full  width  at  half  maximum  (FWHM), ${\rm FOM}=S/{\rm FWHM}$ \citep{Wang2021ultrasensitive,Ng2013Spoof, Cong2015Experimental,Chen2019Toroidal}.  This definition takes into account the variation in the resonant frequency and the width of the resonance that is related to the quality factor by ${\rm FWHM}=f_0/Q$ where $f_0$ is the resonant frequency. Using this definition of the full  width  at  half  maximum, the figure of merit  can be written in terms of the quality factor as ${\rm FOM}=S Q/{f_0}$. From this expression, it seems to be clear that to improve the sensing capabilities the quality factor has to be as large as possible. Nonetheless, it is important to clarify that this reasoning only makes sense for lossless systems where the resonances create maximum variations in the transmission/reflection spectrum.  In the presence of losses, this definition of the FOM can be misleading. As we have demonstrated in this work, losses in the structures limit the light-matter interaction and  perturb the spectrum, reducing the maximum variation of  the  transmission/reflection amplitudes.  The result of applying this definition of figure of merit to the structure that we consider in this paper is shown in Table \ref{tab:table1} for different values of $L_{\rm ratio}$. For example, in Figure \ref{RTcompa} and according to this definition \textcolor{black}{the structure with $L_{\rm ratio}=0.92$ has the larger FOM when only taking into account sensitivity and Q factor, but the difference in the transmission minima between this case and the case of $L_{\rm ratio}=0.88$ is very significant.}

It is clear from our analysis that there is  a trade-off  between  the  resonance intensity (defined as the maximum variation in the transmission/reflection spectra) and the  quality  factor.  This behaviour has been also evaluated in other strongly resonant structures where the figure of merit of the resonant structure was defined as ${\rm FOM}=Q  \Delta I$ with $\Delta I$ being the maximum variation in the transmission/reflection spectra \citep{CongFano2015}. However, this definition does not consider performance as refraction index sensors and this  figure of merit  does not include the sensitivity of the structure. For this reason, we propose an alternative definition of the figure of merit to be used for strongly resonant systems. This definition considers sensitivity, quality factor, and  maximum variation in the transmission/reflection spectra and can be written as 
\begin{equation} \label{eq:FOM}
    {\rm FOM}=S \frac{Q}{f_0}\Delta I
\end{equation}
It is important to mention that this FOM accounts for the differences in reflection and transmission spectra. As we can see from Figs. \ref{RTcompa} and \ref{RTcompb}, losses affect differently to the transmission and reflection spectrum, so the performance as a refractive index sensor in transmission and reflection is different. This means that, for the same resonance, this definition of FOM tells us whether measuring using reflection or transmission configuration is more appropriate.

\begin{table}[H]
  \begin{center}
    \caption{Comparison between different figures of merits applied to the transmission spectrum of lossy all-dielectric structures for different values of $L_{\rm ratio}$ }
    \label{tab:table1}
    \begin{tabular}{l|c|c|c} 
      $L_{\rm ratio}$ & $S (\rm THz/RIU)$ &${\rm FOM}=S Q/{f_0}$ & ${\rm FOM}=S \frac{Q}{f_0}\Delta I$ \\
      \hline
      0.88 & 0.046 & 34.87 & 22.46\\
      0.92 & 0.040 & 48.65 & 9.15\\
      0.97 & 0.036 & 48.03 & 1.07\\
    \end{tabular}
  \end{center}
\end{table}

\section{Comparison between quasi-BIC all-dielectric metasurfaces and metallic structures supporting EOT for sensing}

\begin{figure*}
  \centering
\begin{minipage}[]{0.18\textwidth}
    \subfigure[]{\includegraphics[width=1\linewidth]{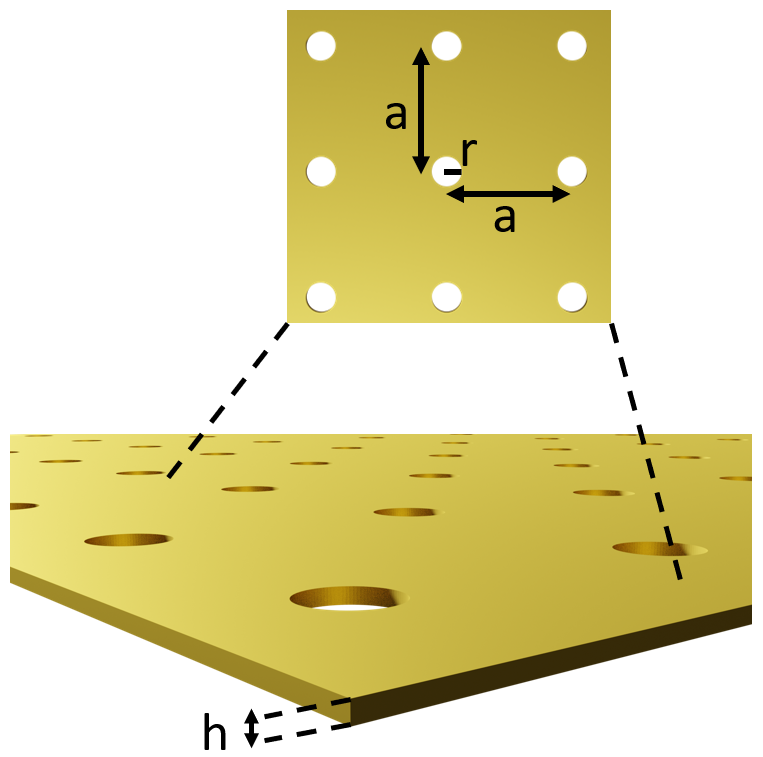} \label{fig:EOTstructa}}
\end{minipage}
\begin{minipage}[]{0.8\textwidth}
   \subfigure[]{\includegraphics[width=0.31\linewidth]{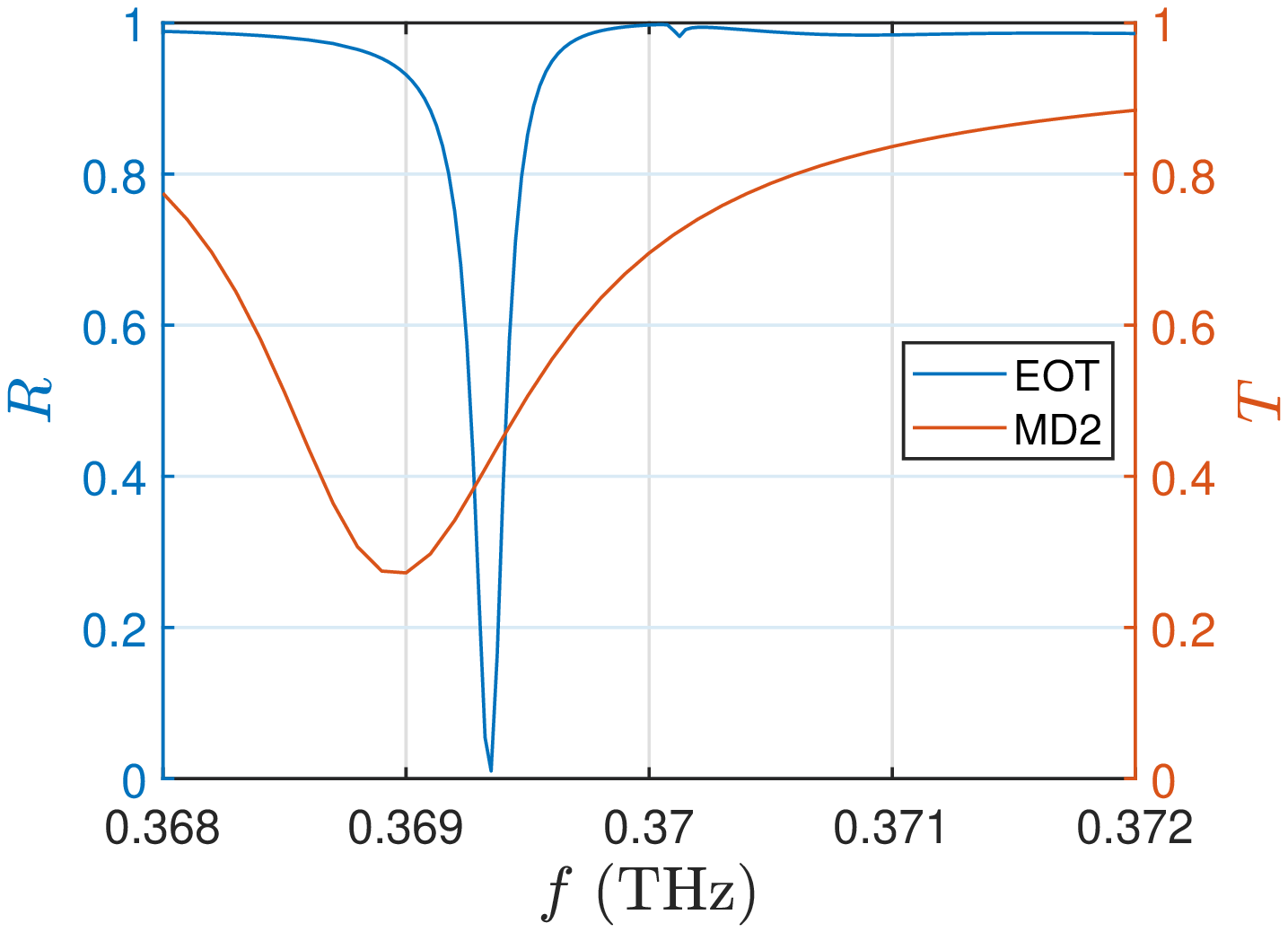}\label{fig:EOTstructb}}
\subfigure[]{\includegraphics[width=0.31\linewidth]{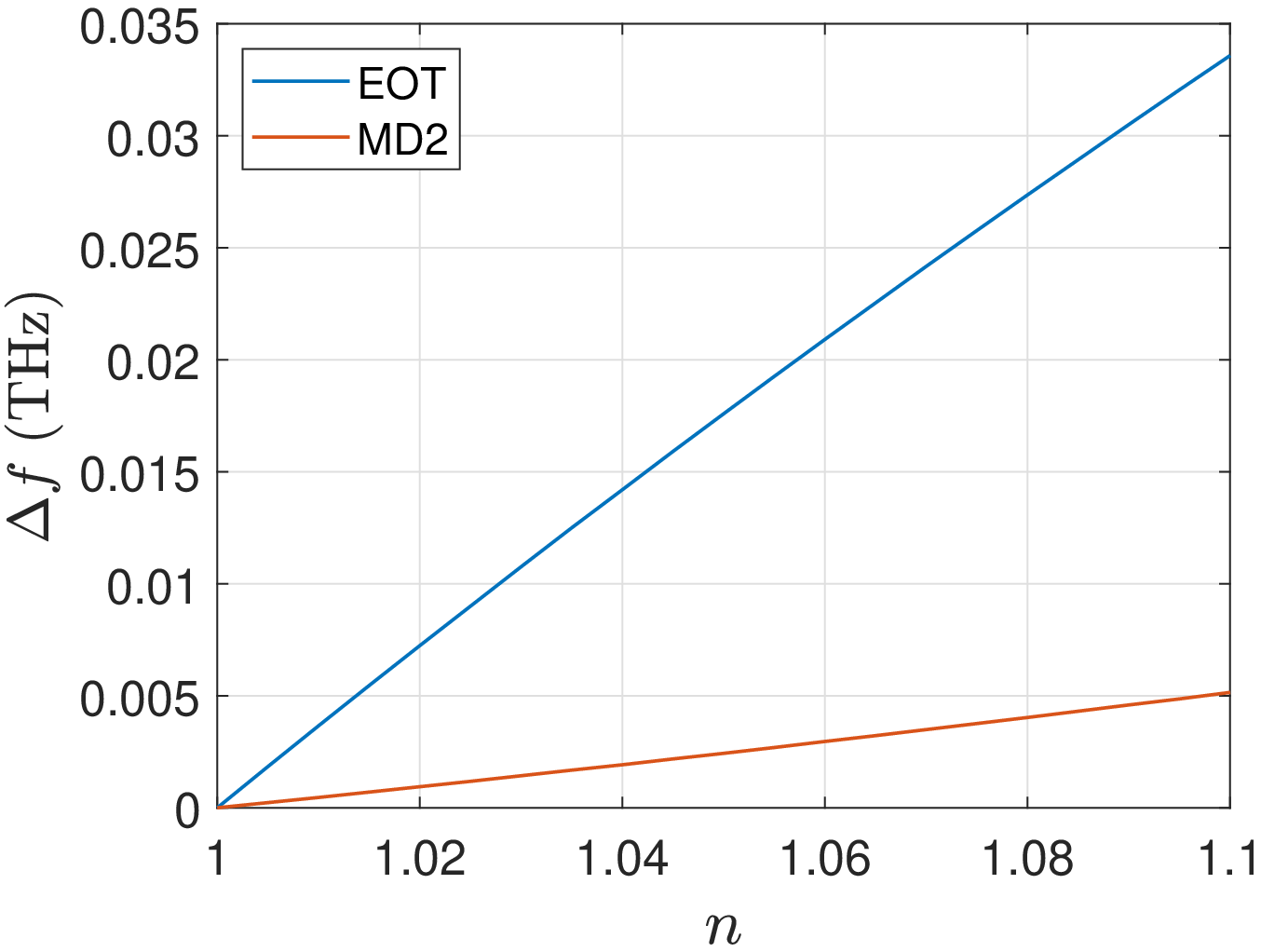} \label{fig:EOTa}}
\subfigure[]{\includegraphics[width=0.31\linewidth]{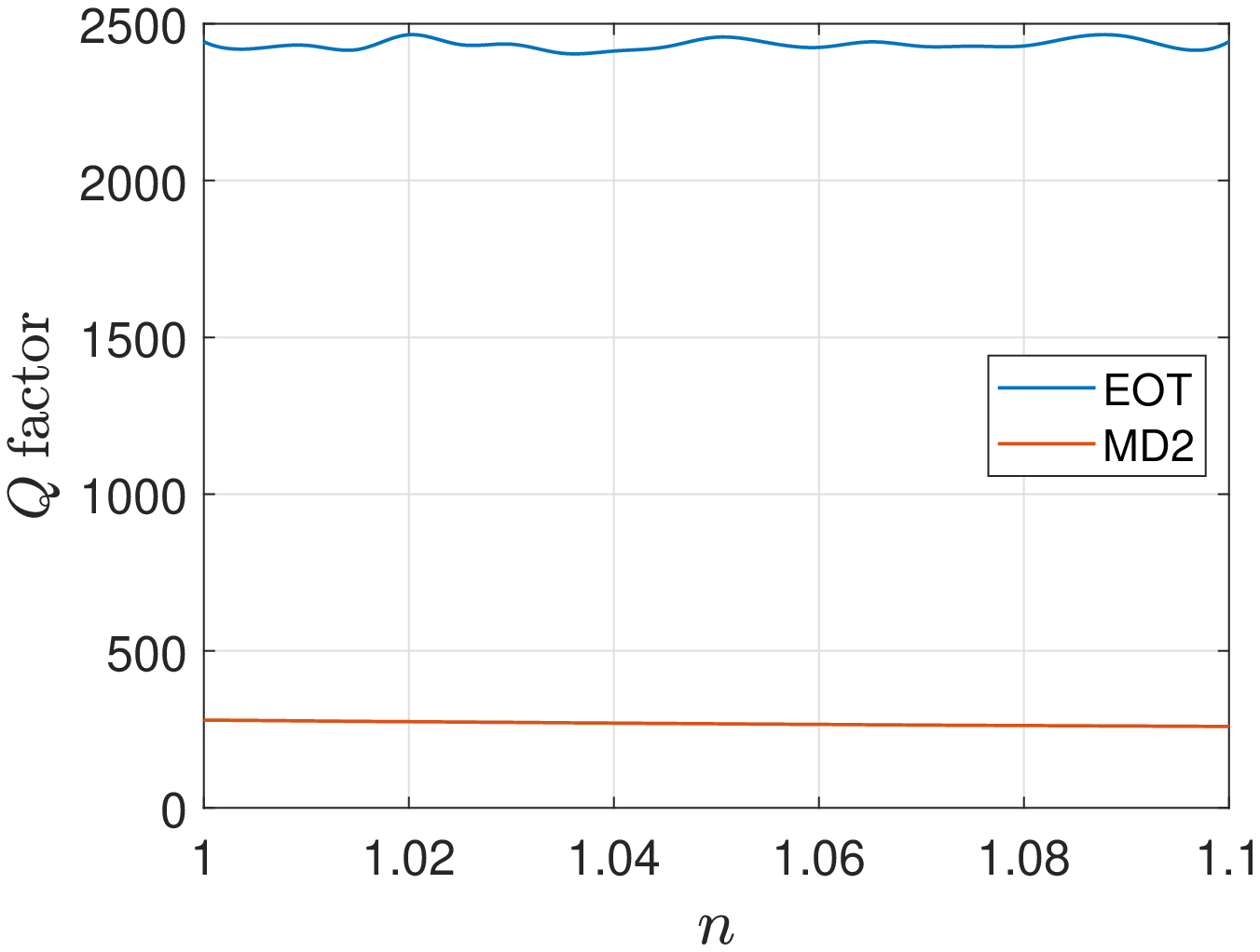}\label{fig:EOTb}}
\subfigure[]{\includegraphics[width=0.31\linewidth]{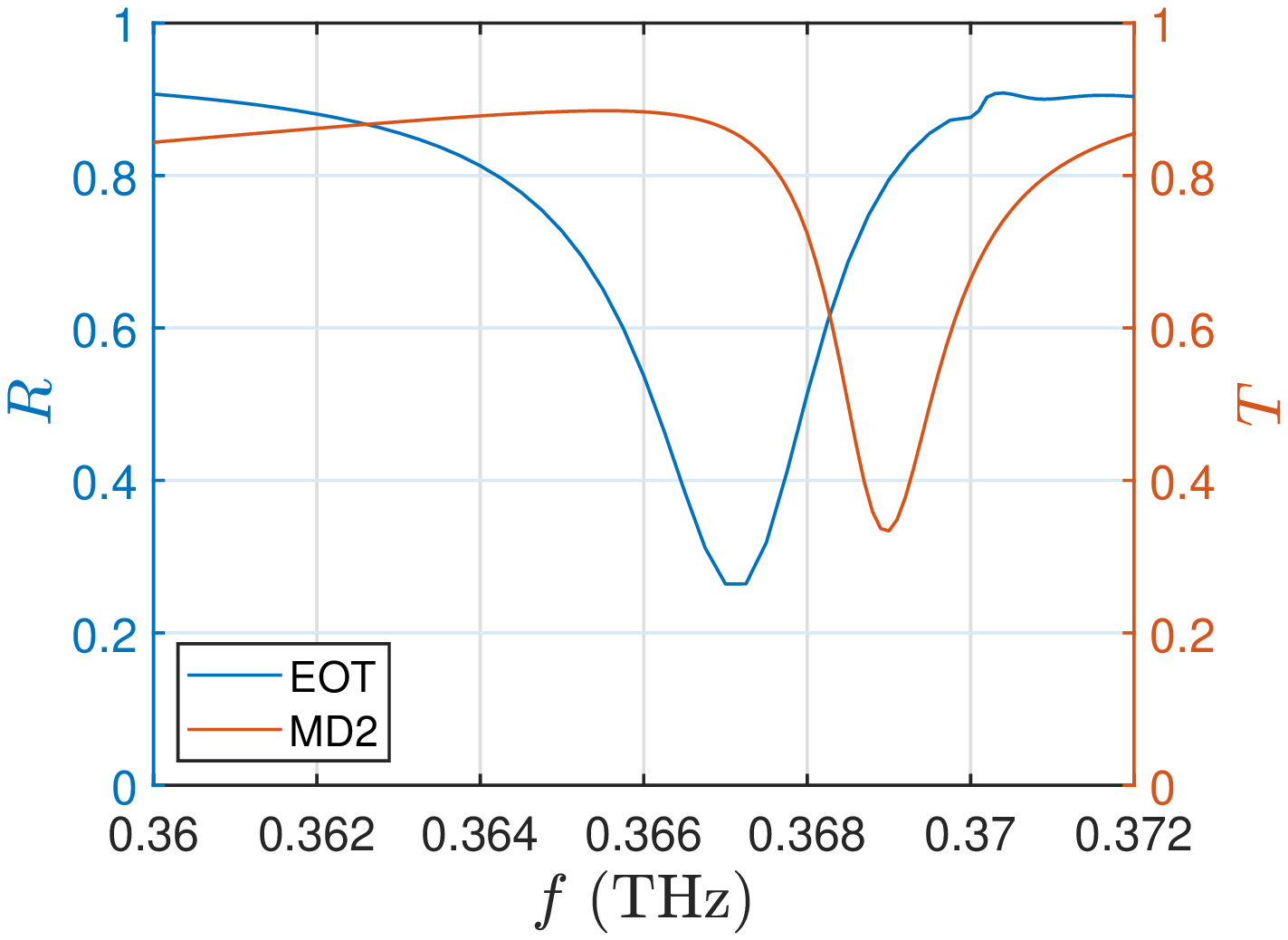} \label{fig:EOTMDa}}
\hspace{0.05in}
\subfigure[]{\includegraphics[width=0.31\linewidth]{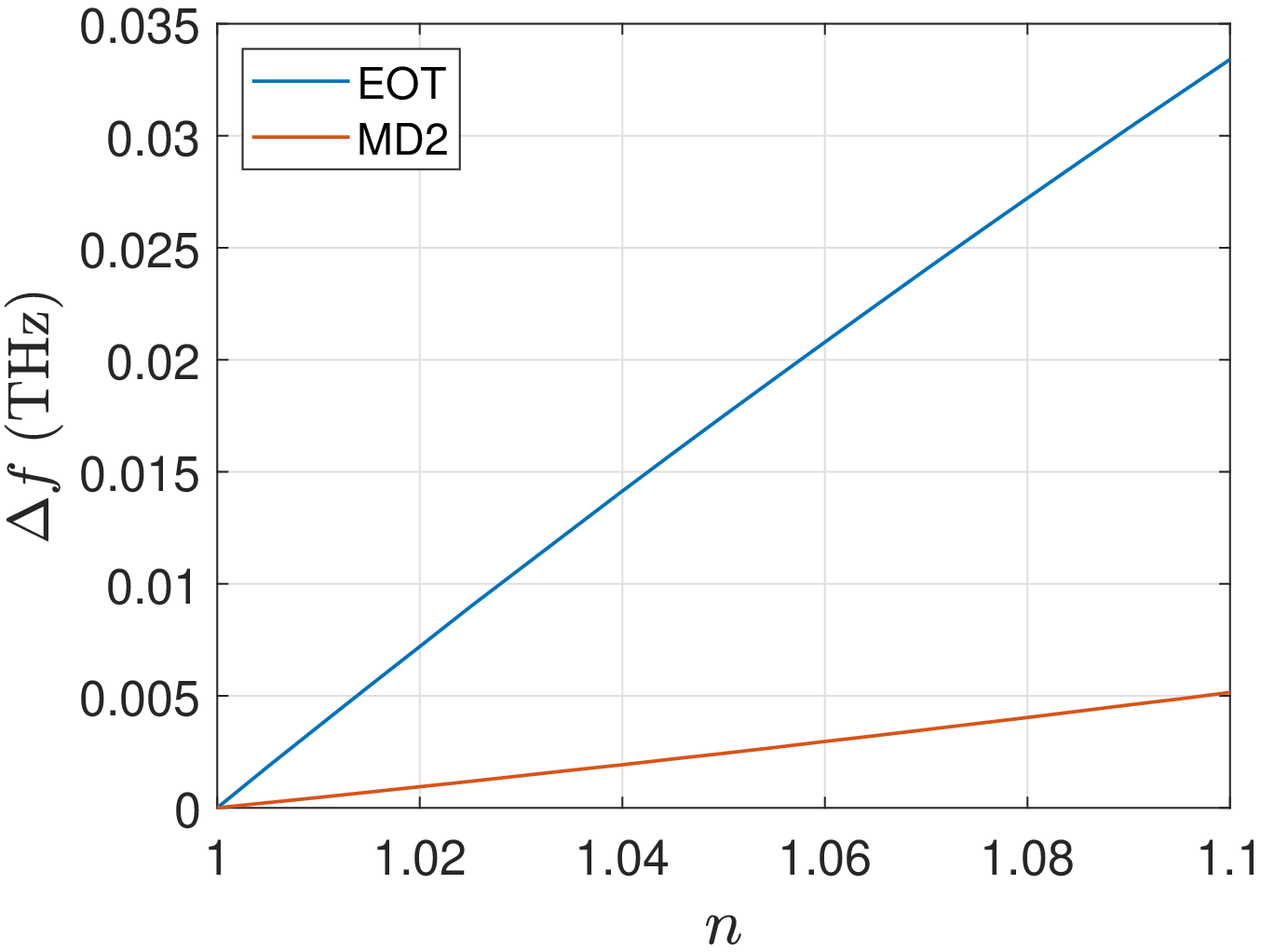} \label{fig:EOTMDb}}
\hspace{0.06in}
    \subfigure[]{\includegraphics[width=0.31\linewidth]{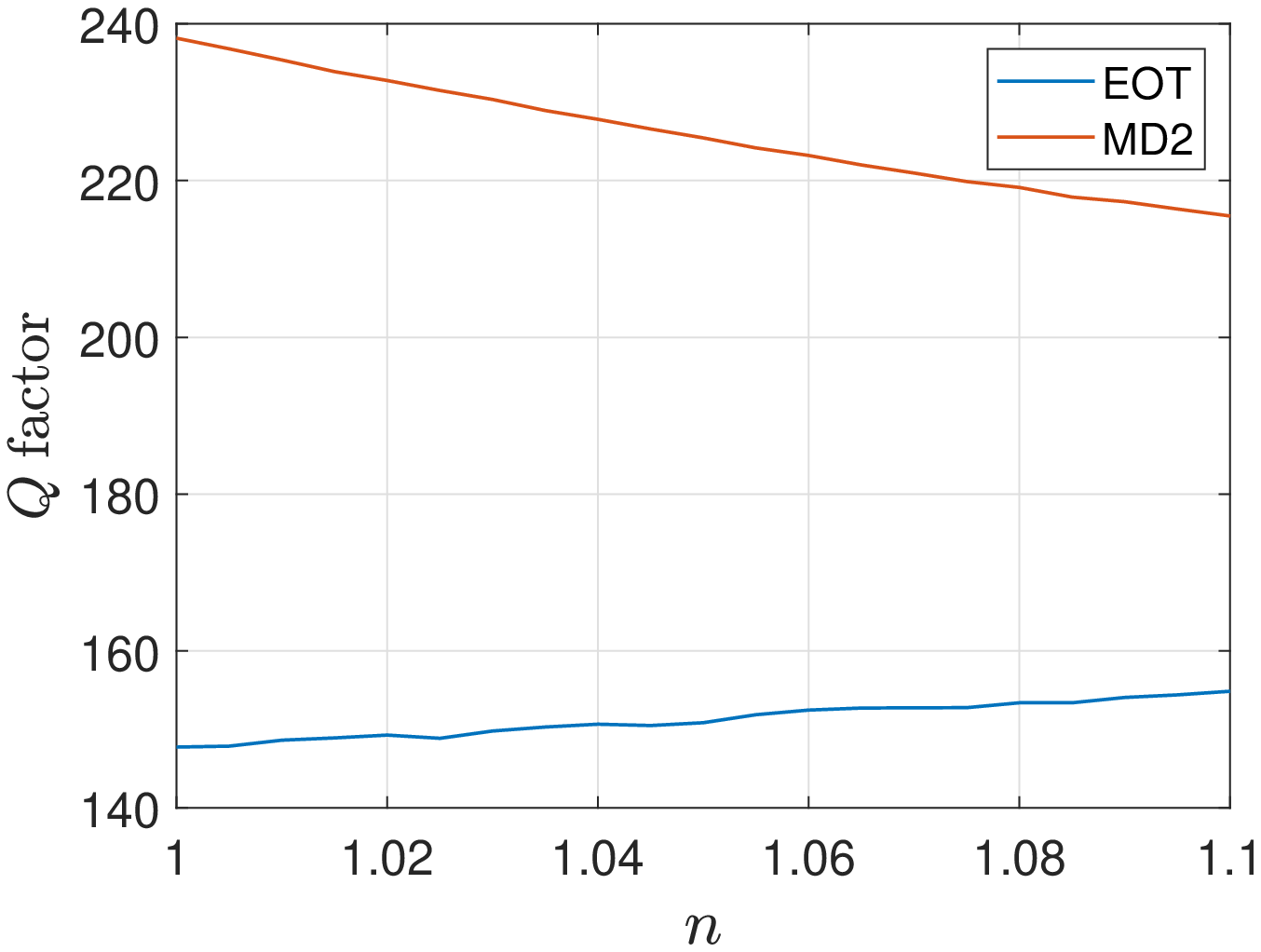} \label{fig:EOTMDc}}
\end{minipage}
  \caption{\textbf{Comparison between quasi-BIC
all-dielectric metasurface and
metallic structures supporting
EOT for sensing}. (a) Schematic representation of a gold plate that supports an EOT resonance. (b)-(d) Comparison of the sensing capabilities with lossless substance under test. The metallic structure with $a=810 \, \mu \rm m$, $h=40 \, \mu \rm m$ and $r=120 \, \mu \rm m$. (b) Transmission spectrum of the dielectric structure and reflection spectrum around the resonance.  Comparison of (c) resonance frequency shift and (d) quality factor as a function of $n$ between EOT and MD2 for $L_{\rm ratio}=0.88$.
  (e)-(g) Comparison of the sensing capabilities with losses in sample. The metallic structure with $a=810 \, \mu \rm m$, $h=50 \, \mu \rm m$ and $r=150 \, \mu \rm m$. Comparison of (e) Frequency spectrum in transmission and reflection for the MD and EOT resonances respectively for refraction index, \textit{n}=1, (f) Resonance frequency shift, (g) quality factor as a function of $n$ between EOT and MD2 BIC for $L_{\rm ratio}=0.89$.}
   \label{fig:EOT}
\end{figure*}

\begin{figure}
  \centering
    \vspace*{-0.1in}
 \subfigure[]{\includegraphics[width=.6\linewidth]{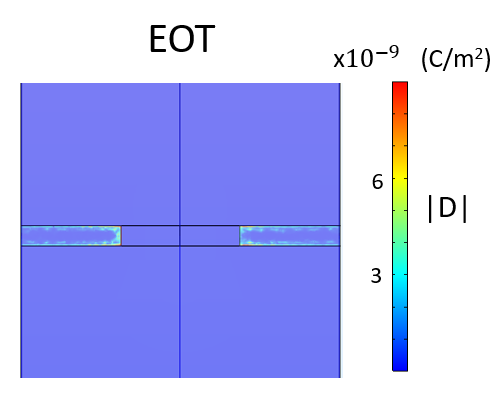} \label{fig:EOTField}}
  \vspace*{-0.1in}
   \subfigure[]{\includegraphics[width=.6\linewidth]{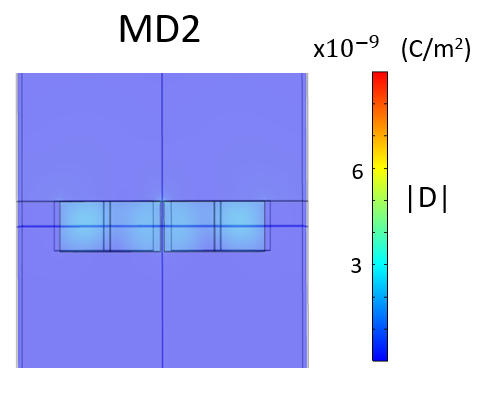}\label{fig:MDField}}
\textcolor{black}{\caption{\textbf{Field profile comparison.} Electric displacement field profile for (a) the EOT resonance, and (b) the MD2 resonance of the dielectric structure.}}
   \label{fig:FieldComp}
\end{figure} 

Finally, we compare the refractive index sensing performance using strong resonances in all-dielectric metasurfaces and the resonance in a perforated metallic layer that supports Extraordinary Optical Transmission (EOT). In this study, we consider two different scenarios. First, the substance under test is considered to be lossless and only the materials of the resonant structures have losses. Secondly, we will repeat the same procedure when the substance under test is lossy.  

We start the analysis of the sensing capabilities with a lossless substance under test by designing a metallic structure that supports EOT. 
Figure \ref{fig:EOTstructa} shows the schematic representation of the structure that consists of a perforated gold plate with a thickness $h=40 \, \mu m$, a distance between holes $a=810 \, \mu m$, and hole radius $r=120 \, \mu m$. 
The metal is modeled using a Drude-Lorentz model that includes losses in the gold layer. \textcolor{black}{This model and its parameters for gold can be found in \citep{Rakic1998Optical}.}
Figure \ref{fig:EOTstructb} shows a comparison of the transmission spectrum for the dielectric structure and the reflection spectrum for the EOT structure. The geometrical parameters are designed to produce a peak of transmission in the same range of frequencies as the dielectric metasurface with $L_{\rm ratio}=0.88$.  Notice that, for the dielectric structure, we consider the response in transmission because it is less affected by losses in the materials [see Figs. \ref{RTcompa} and \ref{RTcompb}]. Following the same reasoning, for the metallic structure supporting EOT, we consider the reflection spectrum that is less affected by losses. From this comparison, we can see that EOT provides a narrower resonance with a larger variation in the response.

To make a comparison between the EOT resonance and the MD resonance in the dielectric structure, we use the quality factor $Q= f _{\rm res}/\rm FWHM$ and the sensitivity to changes in the sample refractive index $S=\Delta f _{\rm res}/\Delta n $. We will assume that a liquid or a gas sample whose refraction index changes depending on the concentration of the substance under test is placed over the structures. In the case of the dielectric metasurface, the presence of the substrate limits the volume where the substance under test can be positioned. We will also consider that the substance under test is thick in comparison with the wavelength at the resonant frequency,  so we can model it as a change in the refraction index of the material above the metasurface. However, in the case of the perforated layer supporting EOT, because the structure is self-supported and it does not need an additional substrate,  the substance under test can be placed on both sides of the structure. 
Figure \ref{fig:EOTa} shows a comparison of the resonance frequency displacement as a function of the sample refractive index for both resonances under these conditions. In this comparison, we can see that the EOT resonance is much more sensitive. From the analysis of the quality factor represented in Fig. \ref{fig:EOTb} we can see that, despite taking into account losses in metal by using the Drude-Lorentz formula, this resonance exhibits a higher quality factor than the MD resonance for $L_{\rm ratio}=0.88$. The results of this study point out that metallic  structures supporting EOT have better performance as sensors with lossless substances.


To complete the analysis we  studied the case of a lossy sample.  We fixed the losses in the substance under test to $k=0.0025$ (the same imaginary part of refractive index as for a quartz substrate). 
We changed the parameters of the EOT structure to $a=810 \, \mu m$, $h=50 \, \mu m$, and $r=150 \, \mu m$ since the resonance with the previous structure parameters had the quality factor too high and the losses in the sample perturbed the response causing disappearance of the resonance. \textcolor{black}{We continue to use $L_{\rm ratio}=0.88$ for the dielectric structure}.
Figure~\ref{fig:EOTMDa} shows the resonances in the two scenarios, and we can see that the intensity of the resonance reaches \textcolor{black}{similar values} in both cases. Figure \ref{fig:EOTMDb} shows that the EOT resonance continues to be more sensitive to the refraction index than the MD2 resonance in the dielectric structure. However, the MD2 resonance has a higher Q factor value, as seen in Fig.~\ref{fig:EOTMDc}. 

\begin{table}[H]
  \begin{center}
    \caption{Comparison between the EOT and the MD resonances using the figure of merit defined in equation (\ref{eq:FOM}) for both the lossless sample case and the sample with losses case.}
    \begin{tabular}{l|c|c} 
      $ $ & EOT FOM & MD FOM\\
      \hline
      Lossless sample  & 2380.23 & 22.46 \\
      \hline
      Sample with losses & 95.07 & 16.63\\
        \end{tabular}
        \label{tab:table2}
  \end{center}
\end{table}

For a better assessment of the sensing capabilities of each structure, we calculate the figure of merit defined by Eq. \ref{eq:FOM}. Table \ref{tab:table2} shows the value of the FOM for both lossy and lossless samples. 
After this analysis, we conclude that the BIC resonances \textcolor{black}{in this dielectric structure have a much worse performance for THz sensing than the EOT resonance, even when comparing the lossless sample case for the MD resonance with the lossy sample case for EOT. 
To study the reason behind this difference in performance, in Fig. 7 we compare the electric displacement field profile (\textbf{D}) in both structures. We can see that the EOT structure reaches higher values of \textbf{D}, but they are confined to an extremely narrow area at the metal surface, with the points of highest field concentration being at the border of the holes. Meanwhile, the MD resonance has its \textbf{D} field concentrated inside the resonators, filling a larger volume, meaning that the absorption will happen over a larger volume and explaining the higher losses compared to the EOT structure. This type of field profile inside the resonators, which is expected for any resonance of an all-dielectric metasurface, combined with the fact that metals behave similarly to perfect electric conductors in the THz band, leads us to the conclusion that dielectric BIC resonances do not present clear advantages for THz sensing over other types of resonances of metallic structures, like EOT.}


\section{Conclusions}
In this work, we have studied the effect of losses in an all-dielectric metasurface that supports multiple resonances in the THz frequency range, some of which come from symmetry-protected BIC that produce strong resonances in the transmission and reflection spectra. 
Although extremely high quality factor values can be obtained through the symmetry breaking when not taking losses into account, absorption in the structure lowers the quality factor and the amplitude of each resonance and imposes a limit on the maximum value of Q obtainable before the resonance stops being visible.
Our analysis shows the effect of losses in the high-index resonators and the substrate separately. 
We find that the substrate is the most limiting element in the structure and that, by making it thinner, it is possible to obtain better values of Q, although the resonance amplitude does not improve compared to the thick substrate case, since the absorption in the cylinders become more important due to the higher quality factor.
The conclusion of our analysis applies to other resonant systems in the presence of losses. For this reason, we propose a figure of merit that provides a better estimation of the sensing capabilities of strongly resonant lossy structures.

Finally, we compare the performance of the all-dielectric structure with a perforated metallic layer supporting EOT resonances. In both the case of a sample without absorption and a sample with absorption, it is possible to design a structure with an EOT resonance that has a higher FOM value that the dielectric structure, meaning that it is more suitable for sensing in this frequency range.

Overall, the main conclusion that can be extracted from this research is that material losses cannot be neglected when studying resonant structures, no matter how low these losses are, especially if the resonances under study have extreme values of the quality factor. Neglecting losses in the design of highly-resonant metasurfaces will lead to errors in the estimation of their scattering properties and the practical applicability of the designs.





\begin{acknowledgments}
This work was funded in part by project PID2019-111339GBI00 and TED2021-132259B-I00 - Spanish Ministerio de Ciencia e Innovaci\'on - Agencia Estatal de Investigaci\'on http://dx.doi.org/10.13039/501100011033 and the Spanish Ministerio de Educaci\'on y Formaci\'on Profesional under the  grant Beatriz Galindo BG20/00024.
\end{acknowledgments}


\end{document}